\documentclass[twocolumn]{aastex631}

\DeclareUnicodeCharacter{2212}{-}

\usepackage{amsmath}
\usepackage{hyperref}

\usepackage{threeparttable}

\usepackage{savesym}
\savesymbol{tablenum}
\usepackage{siunitx}
\restoresymbol{SIX}{tablenum}

\newcommand{\MTRGB}{M_\mathrm{TRGB}}

\definecolor{Deep}{rgb}{0.7, 0.35, 0.7}

\begin{document}


\title{The TRGB-SBF Project. IV.\\ A Color Calibration of the TRGB in the JWST F090W+F150W Filters}


\author[0009-0004-4126-8924]{Maksim I.\ Chazov}
\affiliation{Special Astrophysical Observatory of the Russian Academy of Sciences, Nizhnij Arkhyz, Karachay-Cherkessia 369167, Russia}

\author[0000-0001-9110-3221]{Dmitry I.\ Makarov}
\affiliation{Special Astrophysical Observatory of the Russian Academy of Sciences, Nizhnij Arkhyz, Karachay-Cherkessia 369167, Russia}

\author[0000-0002-9291-1981]{R.\ Brent Tully}
\affiliation{Institute for Astronomy, University of Hawaii, 2680 Woodlawn Drive, Honolulu, HI 96822, USA}

\author[0000-0002-5259-2314]{Gagandeep S.\ Anand}
\affiliation{Space Telescope Science Institute, 3700 San Martin Drive, Baltimore, MD 21218, USA}

\author[0000-0003-0736-7609]{Lidia N.\ Makarova}
\affiliation{Special Astrophysical Observatory of the Russian Academy of Sciences, Nizhnij Arkhyz, Karachay-Cherkessia 369167, Russia}

\author[0000-0001-5487-2494]{Yotam Cohen}
\affiliation{Space Telescope Science Institute, 3700 San Martin Drive, Baltimore, MD 21218, USA}

\author[0000-0002-5213-3548]{John P. Blakeslee}
\affiliation{NSF's NOIRLab, 950 N Cherry Ave, Tucson, AZ 85719, USA}

\author[0000-0003-2072-384X]{Michele Cantiello}
\affiliation{INAF $–$ Astronomical Observatory of Abruzzo, Via Maggini, 64100, Teramo, Italy}

\author[0000-0001-8762-8906]{Joseph B. Jensen}
\affiliation{Utah Valley University, Orem, Utah 84058, USA}

\author[0000-0002-5577-7023]{Gabriella Raimondo}
\affiliation{INAF $–$ Astronomical Observatory of Abruzzo, Via Maggini, 64100, Teramo, Italy}


\begin{abstract}
Observations with JWST in the F090W band provide a powerful tool for determining galaxy distances based on tip of the red giant branch (TRGB) measurements.
It is  a great convenience that the TRGB lies at an almost constant absolute magnitude level at low metallicities.  
However, the TRGB becomes fainter at high metallicities in the F090W filter.
Details of this break in slope are critical for precision applications in the acquisition of distances.
With an absolute scaling set by the maser distance to NGC\,4258 (but excluding the uncertainty in that distance), the value $M^\mathrm{TRGB}_\mathrm{F090W} = -4.40 \pm 0.03$~mag (traditional Vega) is found for $(\mathrm{F090W}-\mathrm{F150W})_0<1.65$~mag.  
The theoretical RGB isochrone that reaches the color 1.65 at the RGB tip corresponds to metallicity $[M/H] = -0.57$ for a 10~Gyr population.
The calibration is used to derive distances for 16 galaxies relative to the megamaser host NGC\,4258. Revised distances are on average slightly closer than literature values derived from the same data.

\end{abstract}




\section{Introduction}\label{sec:intro}

The James Webb Space Telescope (JWST) will enable a measurement of the Hubble Constant through the Population~II ladder of tip of the red giant branch (TRGB) and surface brightness fluctuation (SBF) measurements that is comparable in precision and completely independent of the Cepheid--Type\,Ia supernova ladder~\citep{2001ApJ...553...47F, 2022ApJ...934L...7R}.
The TRGB absolute scale, to be grounded in Gaia parallaxes and RR\,Lyrae cross-referencing \citep{2023A&A...674A..18C, 2024jwst.prop.4783S} will be transferred to SBF, with JWST observations extending to $z\sim 0.06$ enabling the Hubble Constant measurement.

The TRGB calibration is a critical component of the Population~II ladder.
Early observations with JWST confirm theoretical expectations that the absolute magnitude of the TRGB is almost constant at low metallicities in the NIRCam passband 
F090W \citep{2024ApJ...975..195N, 2024ApJ...966...89A, 2024ApJ...973...83A, 2025ApJ...982...26A}.
However, it demonstrably becomes fainter at high metallicities, and there might be a slight dependency at low metallicities as well.
These issues must be understood for the TRGB method to reach its full potential accuracy.
The intent of this study is to determine the precise metallicity dependence of the F090W TRGB absolute calibration.

\subsection{Observed and Theoretical TRGB Color Dependency at F090W}

\begin{figure}
    \centering
    \includegraphics[width=0.9\linewidth]{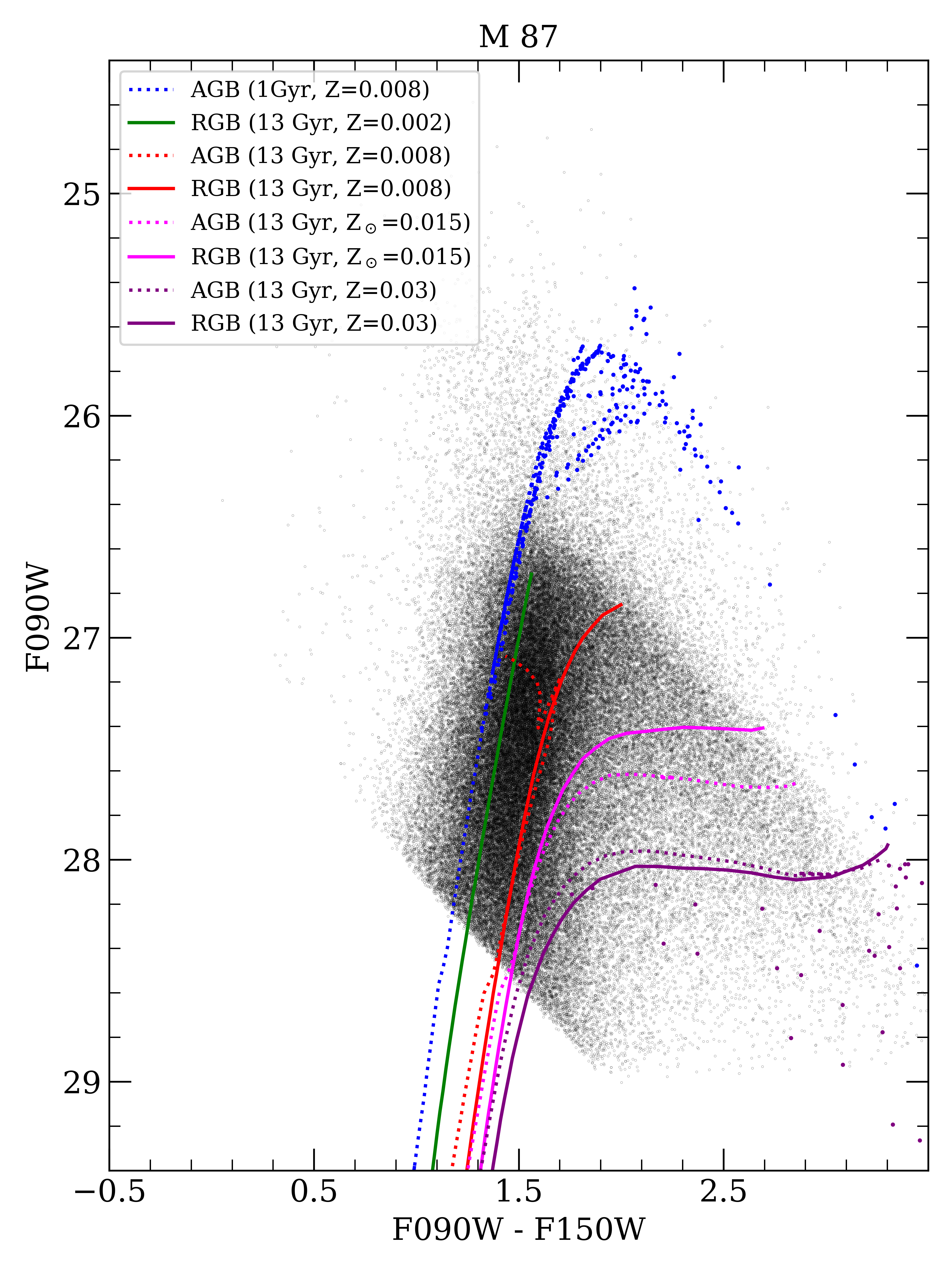}
    \caption{CMD of M\,87 with the PARSEC isochrones overplotted. The isochrones were shifted to $D=16.3$~Mpc \citep{2025ApJ...982...26A} and Galactic extinction reproduced from \citet{2011ApJ...737..103S}. Figure extracted from \cite{2025ApJ...982...26A}.}
    \label{fig:cmd_iso}
\end{figure}

\begin{figure}
    \centering
    \includegraphics[width=0.9\linewidth]{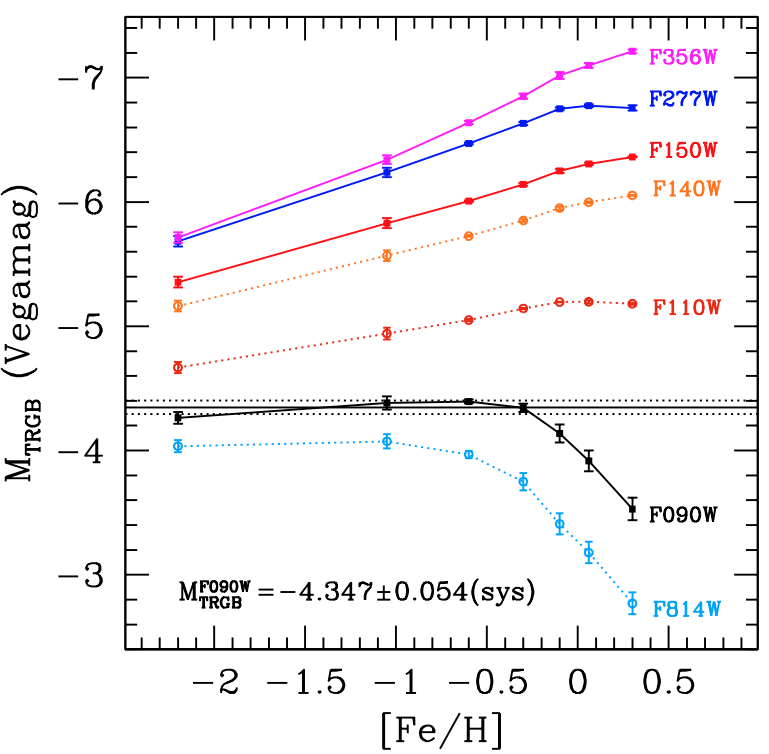}
    \caption{Teramo SPoT model TRGB magnitudes as a function of metallicity for selected JWST and HST filters. Horizontal black line illustrates the F090W calibration from \citet{2024ApJ...966...89A} with one-sigma uncertainties shown with dotted lines. 
    }
    \label{fig:filters}
\end{figure}

The $\mathrm{F090W}_0$ versus $(\mathrm{F090W}-\mathrm{F150W})_0$ color-magnitude diagram (CMD) for stars in the halo of the central galaxy in the Virgo Cluster, M\,87, is shown in the Figure~\ref{fig:cmd_iso} that is drawn from \citet{2025ApJ...982...26A}.
Though asymptotic giant branch (AGB) stars at higher luminosities are a contaminant, the red giant branch (RGB) stars dominate and the trend of the upper envelope to fainter magnitudes at redder colors is evident.
Selected 13 Gyr PARSEC isochrones \citep{2012MNRAS.427..127B, 2013MNRAS.434..488M} give a reasonable representation of the RGB stars from theoretical models.
The expectation from models is further illustrated in Figure~\ref{fig:filters} which plots model magnitudes using the Teramo SPoT stellar population code \citep{2000A&AS..146...91B, 2009ApJ...700.1247R}, showing TRGB values as a function of metallicity for various JWST and HST filters.
It is seen that, in the F090W band, the TRGB absolute magnitude $\MTRGB$ lies at a roughly constant magnitude at low metallicities, only breaking to fainter magnitudes around $[\mathrm{Fe}/\mathrm{H}]\simeq -0.3$. 
The solid horizontal line in the plot is at the absolute value $\MTRGB^\mathrm{F090W} = -4.347$~mag (Vega) determined by \citet{2024ApJ...966...89A} based on the maser distance to NGC\,4258. 
Likewise, there is rough constancy of the TRGB at low metallicities in the HST F814W band~\citep{2021ApJ...915...34H, 2024ApJ...966..175N}, breaking downward at somewhat lower metallicity than at JWST F090W.
At all bands longward of 1~$\mu$m the TRGB increases in magnitude with increasing metallicity requiring precise knowledge of color dependencies for useful distance measurements \citep{2018SSRv..214..113B}.

Constancy of the TRGB at low metallicities and intermediate to old ages was equally found in synthetic photometry using the PARSEC stellar library by \citet{2019ApJ...880...63M}. 
This result was confirmed with JWST data by \citet{2024ApJ...975..195N} who gave attention to the properties of the TRGB in combinations of six JWST passbands based on observations of six nearby galaxies: WLM, Sextans\,A, M\,81, NGC\,253, NGC\,300, and NGC\,2403 with JWST programs ERS--1334 \citep{2017jwst.prop.1334W}, GO--1619 \citep{2021jwst.prop.1619B}, and GO--1638 \citep{2021jwst.prop.1638M}.
The TRGB coverage is robust in the cases NGC\,253, NGC\,2403, and WLM giving good coverage of a wide range of metallicity/age conditions at sub-solar metallicities.
Galaxy distances were taken from HST F814W TRGB measurements \citep{2024ApJ...966..175N}.
From the combination of these data sets, \citet{2024ApJ...975..195N} found the TRGB magnitude at F090W to be constant within the uncertainties over the color range $1.15< (\mathrm{F090W}-\mathrm{F150W})_0 <1.68$.

\subsection{F090W Observations Extending to High Metallicities}

The CMD seen in Figure~\ref{fig:cmd_iso} is an example drawn from JWST GO--3055 that targeted 14 massive early-type galaxies observed to establish a scale link between TRGB and SBF.
Results for 13 of these galaxies have been published \citep{2024ApJ...973...83A, 2025ApJ...982...26A}.
In all cases, the TRGB measurements are made in halo fields where the stars are expected to be old but with a wide range of metallicities.
All 14 have $\mathrm{F090W}_0$ versus $(\mathrm{F090W}-\mathrm{F150W})_0$ CMDs that are remarkably similar in morphology reflecting similar metallicity conditions.  
For 3 cases in the Fornax Cluster see Fig.~3 in \citet{2024ApJ...973...83A} and for 10 cases in and near the Virgo Cluster see Fig.~5 in \citet{2025ApJ...982...26A}.
It is a testament to the utility of the TRGB methodology that these galaxies can be ordered in distance by eye, from the nearest M\,105 to the furthest NGC\,1399.
The collection of individual fits as a function of color to the CMD of these 14 galaxies, to be described in Section~\ref{sec:colorrel}, will provide an empirical formulation of the $(\mathrm{F090W}-\mathrm{F150W})_0$ TRGB--color dependency at high metallicities.  

The pivotal galaxy in this discussion is NGC\,4258.  
This galaxy will be used here to set the absolute scale, assuming the geometric maser distance modulus of $\mu = 29.397\pm0.032$~mag ($D=7.576\pm0.11$~Mpc; \citealt{2019ApJ...886L..27R}).
Multiple fields have been observed at F090W and F150W bands in NGC\,4258 (GO--1685, GO--2875) because of the galaxy's importance for the parallel effort to establish the JWST version of the Cepheid period-luminosity relation \citep{2023ApJ...956L..18R}.
Using data from these programs, \citet{2024ApJ...966...89A} applied the Maximum Likelihood Estimator (MLE) method~\citep{2006AJ....132.2729M} to determine the absolute magnitude of the TRGB at F090W, assuming a constant value with metallicity, of $\MTRGB^{F090W} = -4.347\pm0.033~(\rm{stat.})\pm0.054~(\rm{sys.})$~mag (Vega).
The observations that will be analyzed are of fields in the halo of NGC\,4258.
Stars of a substantial range in age and metallicity can be anticipated to inhabit the halo of this massive spiral galaxy.

\subsection{Metallicity Effects Along the RGB}

The RGB provides a valuable diagnostic tool for studying different stellar populations within galaxies. Specifically, the colors of stars with varying metallicities across the RGB distinguish between different populations. This claim relies on evidence from both observational and theoretical studies, which consistently demonstrate the clear separation of RGB stars based on metallicity \citep{2004AJ....127.2133R, 2012MNRAS.427..127B}.

There is an age-metallicity degeneracy: not only are higher metallicity stars redder but also older RGB stars are redder. We expect that the consequence of this degeneracy will be small in our case. 
Even in the early study by \citet{1993ApJ...417..553L}, it was demonstrated that age variations exceeding several Gyr in old RGB populations result in a significantly smaller color range than variations in metallicity at the the same age. 
Most of our observational sample derives from the extended halos of massive early-type galaxies, so we can reasonably expect that the  resolved stellar populations are highly dominated by old red giants.

\begin{figure}
    \centering
    \includegraphics[width=0.95\linewidth]{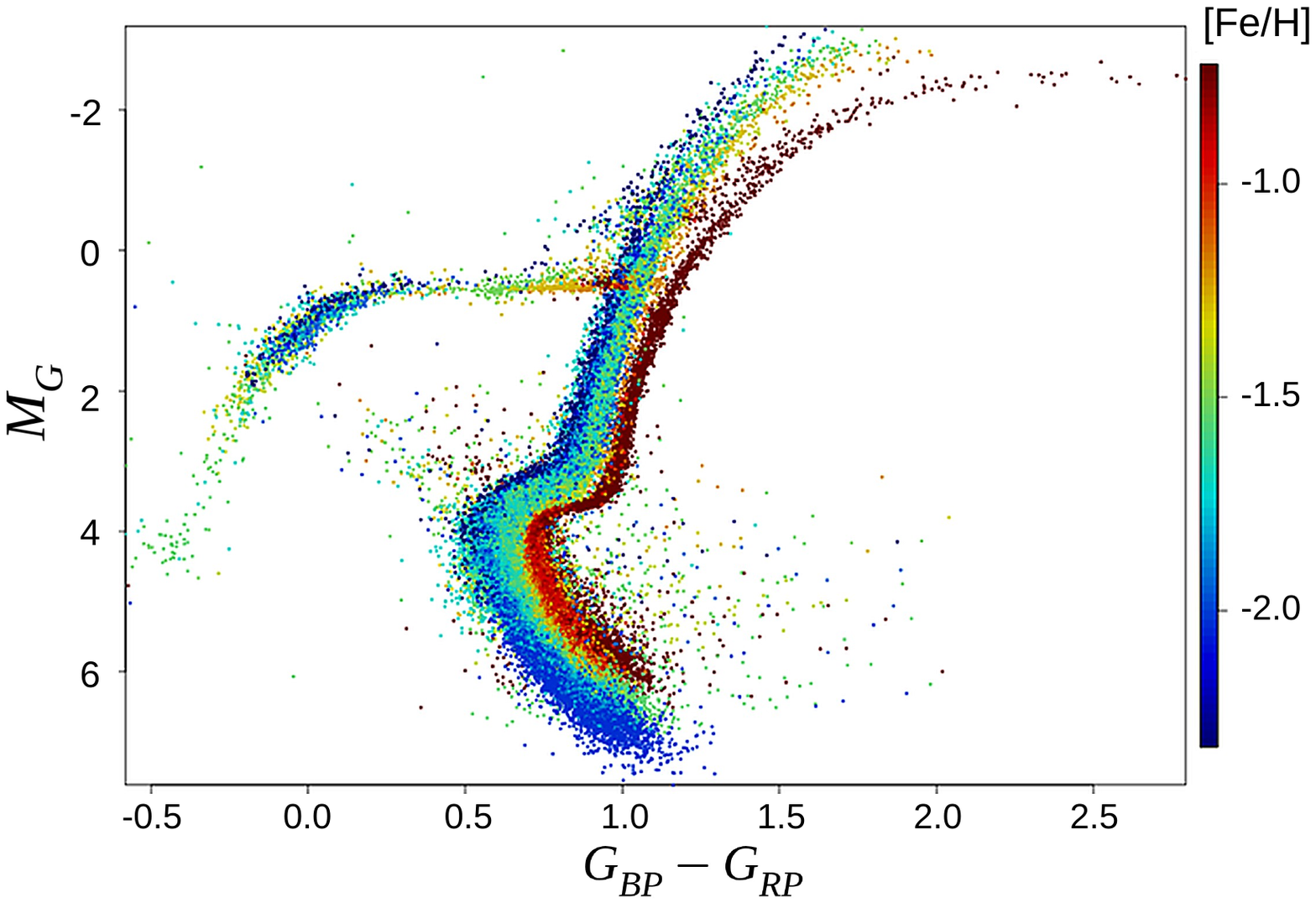}
    \includegraphics[width=1\linewidth]{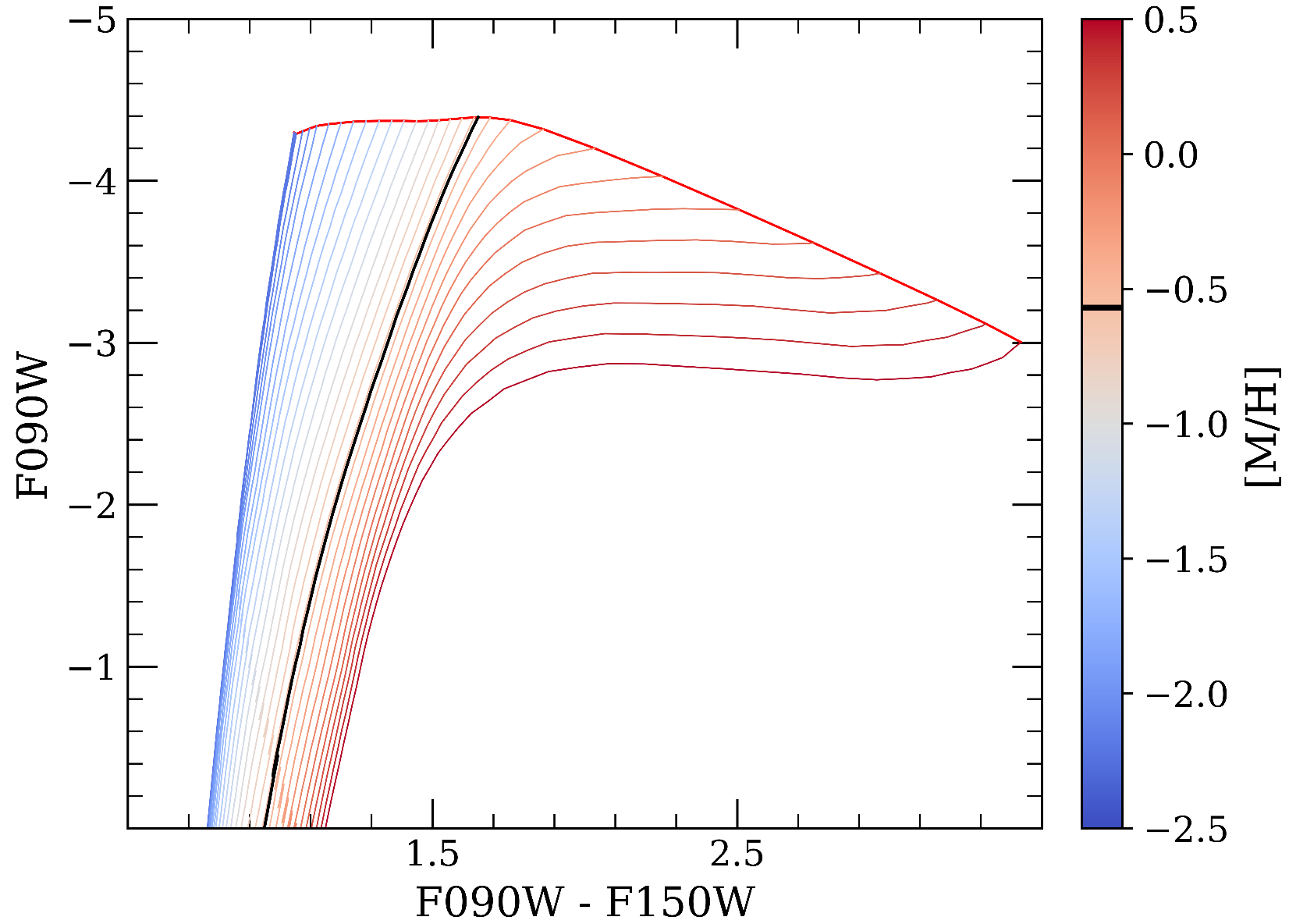}
    \caption{Top panel: CMD of globular clusters observed with Gaia~(Fig.~3 in~\citealt{2018A&A...616A..10G}). Colors from blue to red distinguish increasing metallicities.; Bottom panel: theoretical RGB isochrones from \textsc{padova}~\citep{2012MNRAS.427..127B} for $t=10$~Gyr and $[M/H]= -2.5$ to 0.5. The isochrone for $[M/H]= -0.57$ (TRGB color at $(\mathrm{F090W}-\mathrm{F150W})_0 = 1.65$) is shown by the black line.}
    \label{fig:slicing-clarification}
\end{figure}

Observations of globular clusters using data from the Gaia mission have provided a wealth of information on stellar populations \citep{2026A&A...706A..87L}. The top panel of Figure~\ref{fig:slicing-clarification} presents the CMDs of clusters containing distinct populations demonstrating that RGB stars associated with different populations occupy distinct locations in the CMD. The branches corresponding to different populations are non-overlapping in the RGB region, which is particularly significant because it confirms that metallicity is the dominant factor influencing the position of a star along the RGB. This direct correlation between metallicity and RGB color provides a powerful means of distinguishing populations. More metal-rich stars are typically found to have redder colors, creating a well-defined and observable separation.

Theoretical stellar isochrones reinforce these observational findings, providing a robust framework for understanding the relationship between metallicity and the RGB. The bottom panel of Figure~\ref{fig:slicing-clarification} shows that isochrones calculated for old RGB populations with different metallicities consistently show that RGB tracks remain separate and do not intersect, mirroring what is observed in empirical data.

Together, the agreement between observational evidence and theoretical predictions provides a strong foundation for using RGB stars with varying metallicities as a method to separate and identify stellar populations. By analyzing ``slices'' along the RGB, specific metallicity and age populations can be isolated. This technique will be used in the current work to identify TRGB values as a function of color across the RGB branch.

\smallskip
This paper has two primary goals (a) to establish in detail the color dependency of the TRGB in the JWST F090W passband when coupled with F150W and (b) to provide a recipe for measuring TRGB distances in these filters going forward.
In Section~2 there is discussion of the input material and of characteristics of the RGB. In Section 3 the maximum likelihood procedure for locating the TRGB is described and
then there is focus on the role of NGC\,4258 in the calibration.
Section~4 and the Appendices provide a description of how the CMD is segmented in color intervals to monitor the TRGB with color.
The results of this segmentation analysis are presented in Section~5.
There is application to the current sample of 16 galaxies in Section~6.
Sections 7 and 8 are discussion and conclusion.


\section{Stellar Photometry and TRGB Framework}
\label{sec:framework}

\subsection{Photometry}
\label{sec:photom}

For our analysis, we collected data from the JWST observing programs GO--1638 \citep{2021jwst.prop.1638M}, GO--1685 \citep{2021jwst.prop.1685R}, and GO--3055 \citep{2023jwst.prop.3055T}. The data were reduced with the identical calibration pipeline (\texttt{CAL\_VER}=1.17.1), reference file (\texttt{pmap}=1322), and DOLPHOT/NIRCam (February 4, 2024) versions to ensure uniformity across the full set of data. For science that requires extremely high-precision (at the level of $\sim$0.02 mag, or 1$\%$ in distance), we argue that such a level of care is required. For instance, in the case of NGC\,4258, changes to both the JWST/NIRCam pipelines (including updates to underlying reference files)  and DOLPHOT since \cite{2024ApJ...966...89A} result in a systematic increase in the brightness of the TRGB feature at the level of $\sim$0.03~mag, which was determined by using the same measurement procedures in that paper on the newly reduced set of photometry presented here.

Stellar point-spread function (PSF) photometry of images was carried out using the NIRCam module of \textsc{dolphot}~\citep{2016ascl.soft08013D, 2024ApJS..271...47W} following the recommendations laid out within \cite{2024ApJS..271...47W}. We adopted the following criteria \citep{2023RNAAS...7...23W, 2023ApJ...956L..18R} for selection of photometric sources: (1)~$\text{Crowding}<0.5$, (2)~$(\text{Sharpness})^2\leq0.01$, (3)~$\text{Object Type}\leq2$, (4)~$S/N\geq5$, and (5)~$\text{Error Flag}\leq2$. As was shown in \citet{2025ApJ...982...26A}, there was no significant systematical shift between TRGB values derived from different NIRCam detectors. For analysis performed in this study we used the combined photometry from all chips. Similarly, for NGC\,4258 and NGC\,5643 from GO--1685, we combined photometry from two separate (non-overlapping) visits.

Photometric uncertainties, bias, and completeness are evaluated by the injection and recovery of synthetic stars matching the spatial, magnitude, and color distribution of the observed stars using \textsc{DOLPHOT's} artificial star mode. In order to take statistical variance into account more carefully and minimize fluctuations we adopted several methods.
The photometric completeness function $\rho(m)$, the bias function $b(m)$, and the photometric error distribution $\varepsilon(m \mid x)$ were estimated using kernel density estimation (KDE) with a Gaussian kernel.
The Median Absolute Deviation (MAD) was used in the calculation of the bias and error functions, allowing for stable and robust approximations.
The resulting discrete functions were interpolated using a univariate spline in the B-spline basis with a smoothing parameter \citep{Virtanen2020}.

To correct for foreground extinction $A_\mathrm{F090W}$ and $A_\mathrm{F150W}$ for each target,
$E(B − V)$ values are taken from~\citet{2011ApJ...737..103S} and the extinctions are found assuming $A_\mathrm{F090W} / E(B − V ) = 1.4156$ and $A_\mathrm{F150W} / E(B − V ) = 0.6021$ ~\citep{2023ApJ...956L..18R}. We have applied these corrections {\it before} the TRGB measurements. 

\subsection{Spatial Selections}
\label{sec:select}

It is favored to make TRGB measurements in the halos of massive galaxies to avoid crowding and contamination from younger stellar populations, as well as to maximize the contrast of low metallicity RGB. As an additional practical matter, highly-precise PSF photometry in crowded fields becomes increasingly difficult due to issues in performing accurate aperture corrections or empirical PSF adjustments. For the 14 early-type galaxies from GO--3055, we performed photometry only on the outermost 2/8 NIRCam short-wavelength detectors to avoid increasing levels of crowding towards the cores of these massive elliptical galaxies (which were centered on the other side of NIRCam for measurements of SBF). 
For examples of these field selections, see \citet{2024ApJ...973...83A} Fig.~2 in the case of NGC\,1380 or \citet{2025ApJ...982...26A} Fig.~4 in the case of M89.

For galaxies with a component of their populations in young stars (e.g., NGC\,4258), a quantitative choice is to exclude parts of the galaxy with surface brightness greater than the photographic blue isophote 25~mag\,arcsec$^{-2}$ \citep{2021MNRAS.501.3621A}. As a test of this cutoff magnitude, we explored the use of stellar photometry at higher surface brightnesses in the case of NGC\,4258. NGC\,253 aside, this galaxy is the nearest of our calibration sample by a factor 2 to 3, providing better spatial resolution at given surface brightness.  The particular interest was to access higher metallicity RGB stars than available in the outer halo.  As seen from the footprints of the repeated observations of NGC\,4258 in \citet{2024ApJ...966...89A} there is a wide radial range of coverage in this galaxy.

\begin{figure}
\centering
\includegraphics[width=1\linewidth]{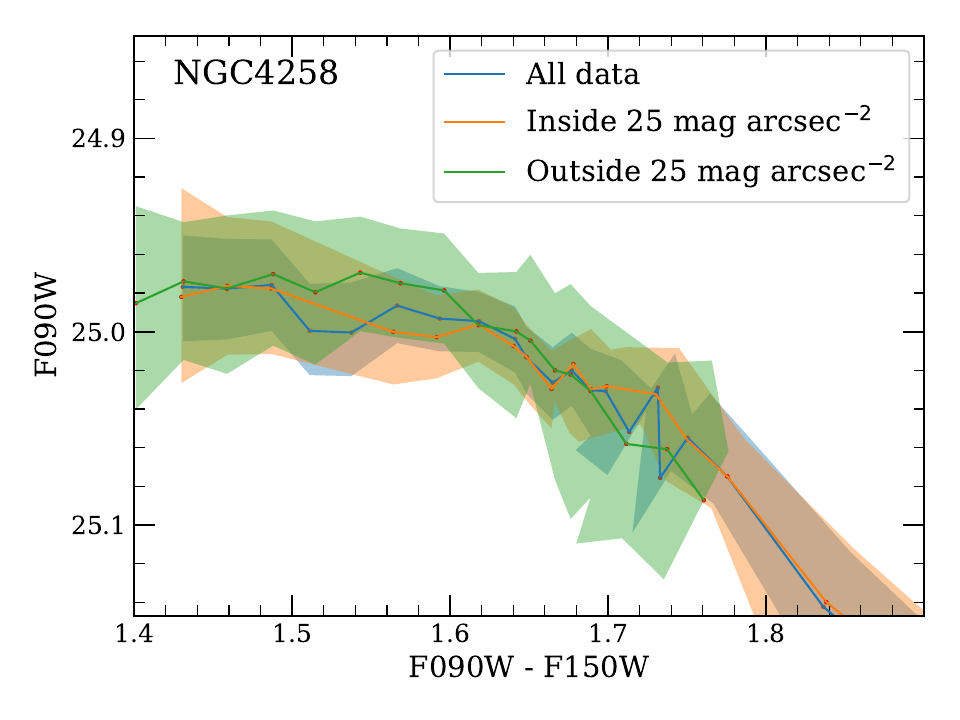}
\caption{NGC\,4258: TRGB color relation comparison for stellar populations outside 25~mag\,arcsec$^{-2}$ isophote and inside this isophote to a radius 0.6 times the radius at 25~mag\,arcsec$^{-2}$.}
\label{fig:NGC4258:in-out}
\end{figure}

Certainly, crowding is extreme at inner radii, making photometric measurements of individual stars impossible.  From experimentation, it was concluded that NGC\,4258 CMD were usefully uncrowded outside 0.6 times the 25~mag\,arcsec$^{-2}$ radius.
With Figure~\ref{fig:NGC4258:in-out} there is a comparison of the TRGB dependence on color (with the measurement techniques described in Section~\ref{sec:segmentation}) at radii outside 25~mag\,arcsec$^{-2}$ and the useful domain inside that isophote. 
The TRGB color relations do not show any statistically significant difference between the inner and outer regions and are well described by the color relation obtained for the combined PSF--photometry. 
Nonetheless, we proceed with only measurements from the outer regions to ensure the highest-fidelity photometry.

\section{Maximum Likelihood Estimation (MLE)}
\label{sec:mle}

The maximum likelihood methodology for determining TRGB distances was proposed by \citet{2002AJ....124..213M} and further developed by \citet{2006AJ....132.2729M}. 
In this approach the true luminosity function (LF) is described by split power laws for RGB and AGB components and a jump at the TRGB.
\begin{equation}
\psi = \begin{cases}
  10^{\,a(m-m_\mathrm{TRGB})+b}, & \mbox{if } m \geq m_\mathrm{TRGB} \\ 
  10^{\,c(m-m_\mathrm{TRGB})},   & \mbox{if } m < m_\mathrm{TRGB}.
\end{cases}
\label{eq:RGBLF}
\end{equation}
Here, $m_\mathrm{TRGB}$ is the magnitude of the TRGB, $b$ denotes the relative strength of the RGB discontinuity, and $a$ and $c$ are the slopes of the RGB and AGB, respectively.

The analytic description of the RGB LF allows us to compute the observed luminosity distribution in a consistent manner, properly accounting for photometric uncertainties, incompleteness, and other observational effects.
\begin{equation}
\varphi(m) = \int \psi(z) \rho(z) \varepsilon(m|z) \, dz,
\end{equation}
where $\rho(z)$ denotes the photometric completeness function, and $\varepsilon(m|z)$ describes the distribution of photometric errors.
Both functions are derived from artificial-star photometry that mimics the real observations.
As a first approximation for the error function $\varepsilon(m|z)$, it is convenient to adopt a Gaussian distribution whose mean and dispersion depend on the true stellar magnitude.

This formalism allows us to estimate the likelihood function of the observed sample of $N$ stars within the magnitude range $[m_{\min}, m_{\max}]$. 
The probability is given by
\begin{equation}
P(\mathbf{x}) = \prod_{i=1}^{N} \frac{\varphi(m_i|\mathbf{x}) \,dm}{\int_{m_{\min}}^{m_{\max}}\varphi(m|\mathbf{x}) \,dm},
\end{equation}
which we evaluate as a function of the free parameters $\mathbf{x}(m_\mathrm{TRGB}, a, b, c)$.
The corresponding log-likelihood function for minimization is
\begin{equation}
\begin{split}
\mathcal{L}(\mathbf{x}) & = -\log P(\mathbf{x}) \\
                        & =  -\sum_{i=1}^{N} \log \varphi(m_i|\mathbf{x}) + N \log \int_{m_{\min}}^{m_{\max}} \varphi(m|\mathbf{x}) \,dm.
\end{split}
\end{equation}

We have developed a new tool using \textsc{Python}~(Chazov et al., in preparation). This approach is based on the MLE implementation in \textsc{MATLAB} and is an evolution of the previous methodology by \citet{2006AJ....132.2729M}.
It includes both \textsc{Python} scripts and a cross-platform graphical user interface, and will be released in the near future. The key innovation in this new iteration is the ability to perform MLE fits along physically motivated slices of the RGB.

\subsection{Segmentation of the RGB}
\label{sec:segmentation}

\begin{figure}
    \centering
    \includegraphics[width=0.9\linewidth]{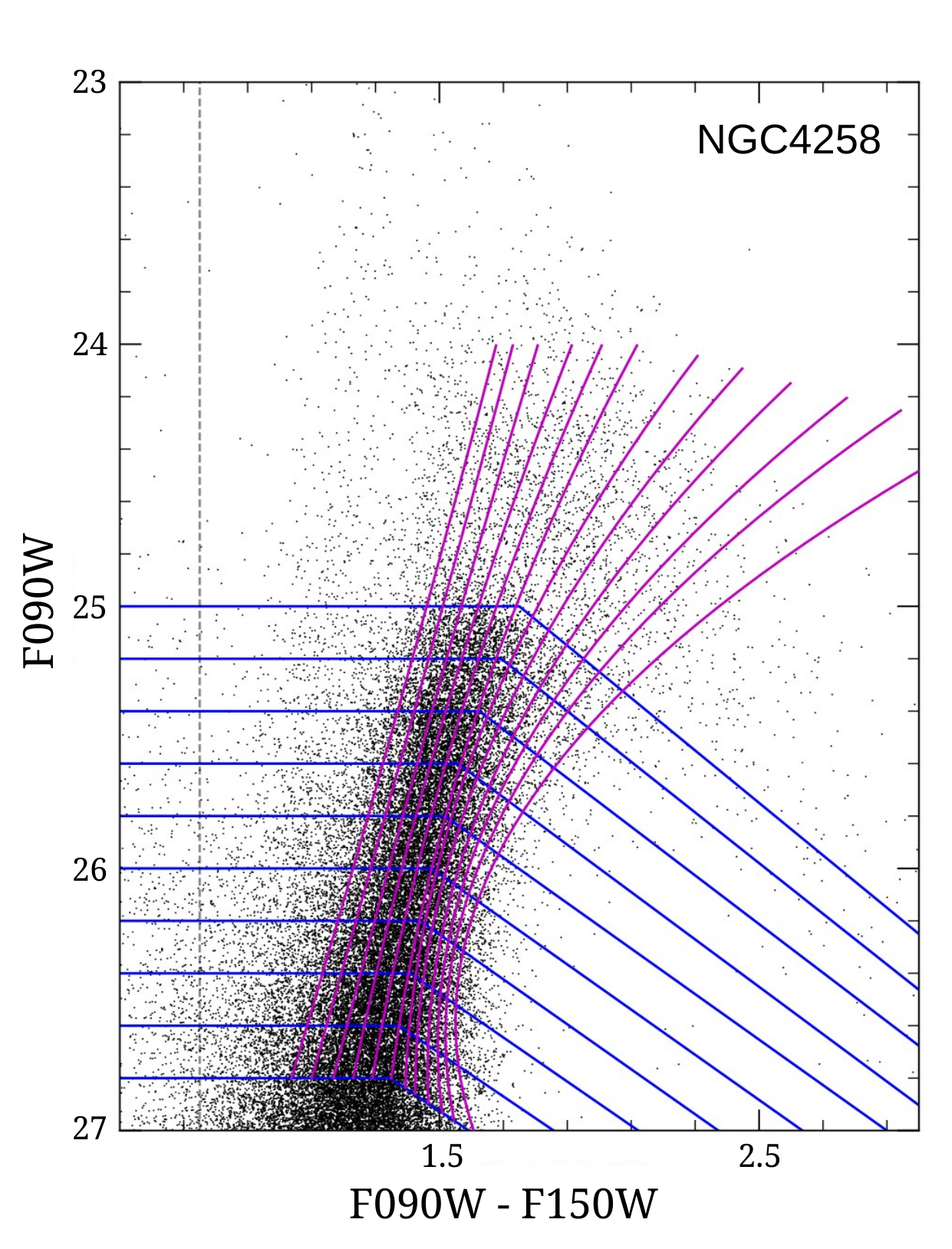}
    \caption{Segmentation of an RGB CMD into color strips bounding individual maximum likelihood tests. Representative magenta lines are shown.  Details discussed in Appendix~\ref{app:RGBsegmentation}.
    }
    \label{fig:strips}
\end{figure}

It is clear that the TRGB in the F090W band gets fainter at redder colors corresponding to higher metallicities.
A quantitative measure of the related color dependence is determined as follows.
First, an RGB, such as the one presented in Figure~\ref{fig:strips}, is segmented into F090W intervals using broken blue lines. In the bluer part of the diagram, these lines are horizontal, whereas in the redder part, they roughly follow the TRGB color relation.
This procedure allows us to split the RGB into evenly spaced strips starting from the tip. 
It enables us to analyze the distribution of stars at fixed distances from the TRGB and to identify distinct populations based on percentiles. 
In the next step, the positions of these percentiles are described by second-order polynomials, which generate RGB slices in the dereddened color $(\mathrm{F090W}-\mathrm{F150W})_0$ space, thereby defining regions that encompass stellar populations spanning a range of ages and metallicities.
This color-based segmentation is described in more detail in Appendix~\ref{app:RGBsegmentation} and is illustrated by the magenta lines in Figure~\ref{fig:strips}.
For each slice, the magnitude distribution is then fitted using the standard broken-exponential RGB LF (Equation~\ref{eq:RGBLF}), and the maximum-likelihood methodology is applied for detecting the tip.
The color assigned to each slice at the location of the fitted TRGB is defined as the mean of the color boundaries of the slice, derived from the fitted magnitude level.

However, the significant TRGB slope for high-metallicity red stars in the F090W filter is a particular challenge for the maximum-likelihood approach.
The slope results in over-smoothing of the observed one-dimensional LF in the vicinity of the TRGB, which in turn degrades its representation by the model and reduces the accuracy of the fit.
To mitigate this effect, we perform TRGB measurements for red stars using the F150W filter, carefully accounting for all its photometric effects. In this filter, the color dependence of the TRGB for high-metallicity stars is substantially weaker and approaches a constant value. This enables higher-precision TRGB determination. A more detailed description of this approach is provided in Appendix~\ref{app:MLfit}.

\begin{figure}
    \centering
    \includegraphics[width=1.0\linewidth]{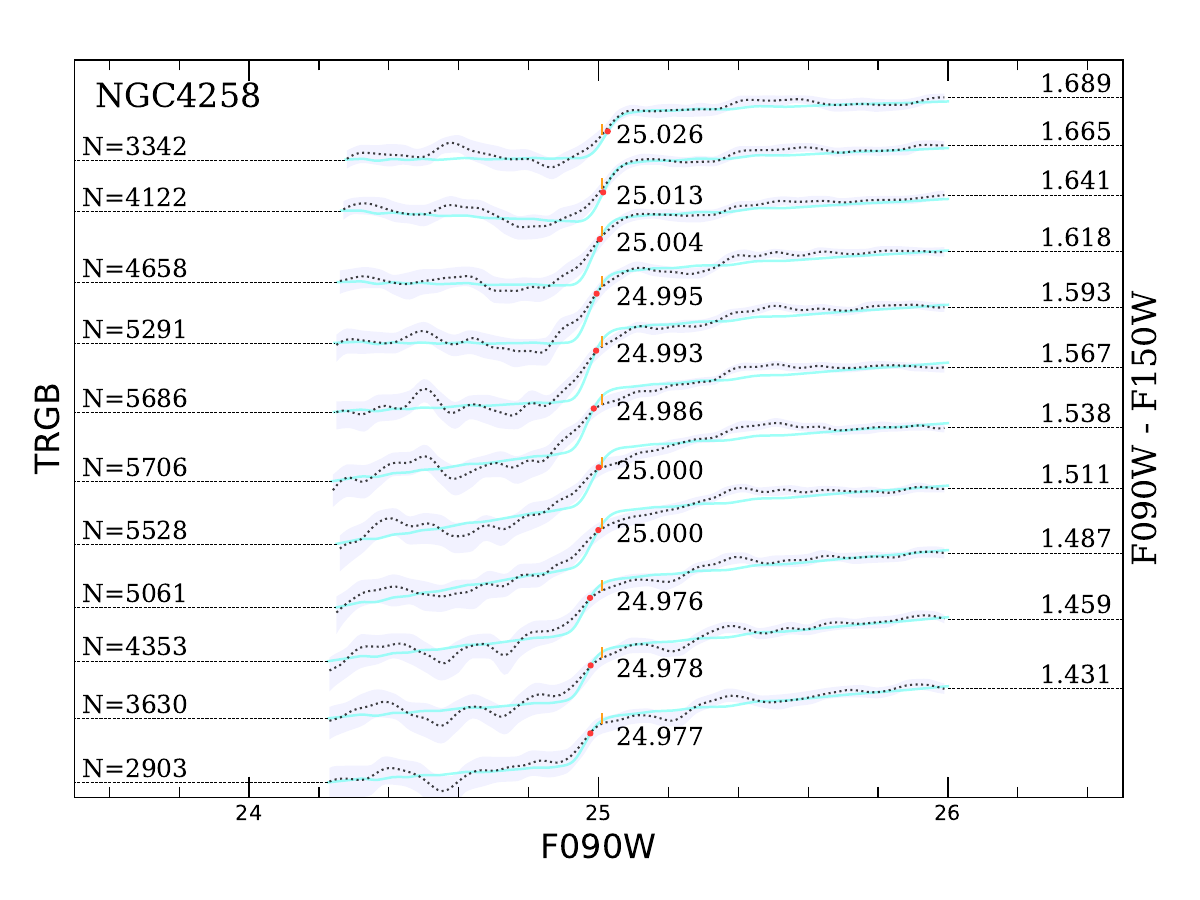}
    \caption{In the case of NGC\,4258, logarithmic RGB LF in color slices from $(\mathrm{F090W}-\mathrm{F150W})_0 = 1.4$ (bottom) to 1.7 (top). The number of stars included in the MLE calculation for each slice is shown on the left. The fitted TRGB magnitudes are displayed in the center numerically and as red dots, while the corresponding slice colors are shown on the right.
    Orange vertical segments at F090W=25.01 record the TRGB values given by \citet{2024ApJ...966...89A}.}
    \label{fig:NGC4258:LF:color}
\end{figure}

In the case of NGC\,4258, Figure~\ref{fig:NGC4258:LF:color} shows the logarithmic LF of the RGB in slices, ranging from the blue extreme at $(\mathrm{F090W}-\mathrm{F150W})_0=1.4$ to the red extreme at $(\mathrm{F090W}-\mathrm{F150W})_0=1.7$. The black dotted lines and the shaded regions correspond to the observed smoothed RGB LF and its errors (95\% confidence level). The cyan broken lines represents the predicted LF, incorporating photometry completeness, errors, and biases. Small orange vertical segments at $\mathrm{F090W}_0 = 25.01$~mag indicate the TRGB values consistent with \citet{2024ApJ...966...89A}, given updated photometry and accounting for reddening.

The revised \citet{2024ApJ...966...89A} result is to be compared with the mean TRGB magnitude of $\mathrm{F090W}_0 = 24.99$~mag for the slices in this figure.
\citet{2024ApJ...966...89A} include higher-metallicity stars in their broader color selection range of $1.30 \leq (\mathrm{F090W}-\mathrm{F150W})_0 \leq 1.75$, which when combined with a modest shift in photometry (see Section 2.1) led to a shift of the TRGB magnitude toward a fainter magnitude. The broader color range selection used in that work was motivated by the predicted behavior of the TRGB in the PARSEC suite of isochrones \citep{2012MNRAS.427..127B}.

\subsection{Thick vs Thin Slices}
\label{sec:thickthin}

\begin{figure}
    \centering
    \includegraphics[width=1.0\linewidth]{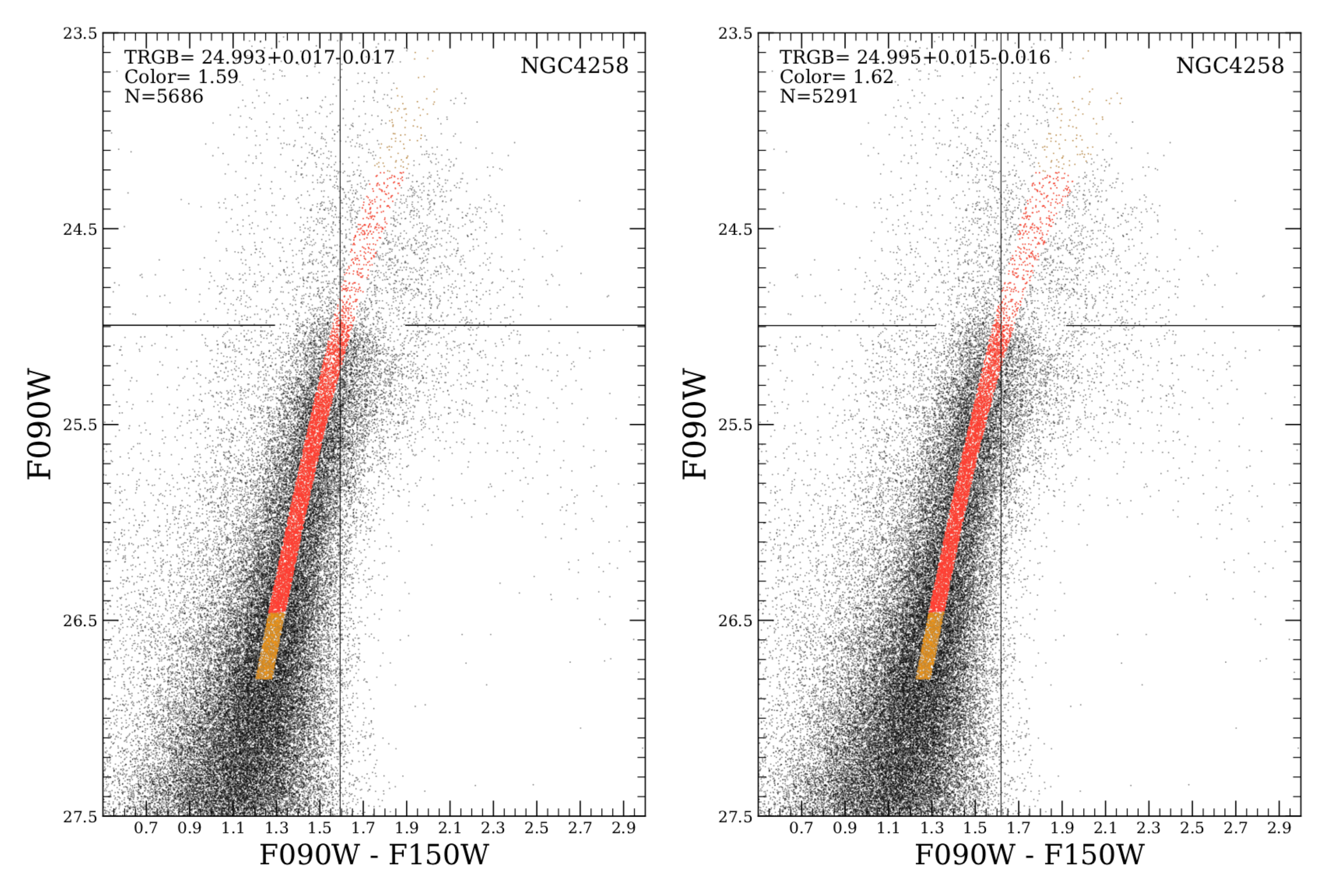}
    \caption{Two representative thin slices of the NGC\,4258 RGB. The slice stars are shown in orange, while the stars used in the MLE calculations are shown in red.}
    \label{fig:NGC4258:slice:thin}
\end{figure}

\begin{figure}
    \centering
    \includegraphics[width=1.0\linewidth]{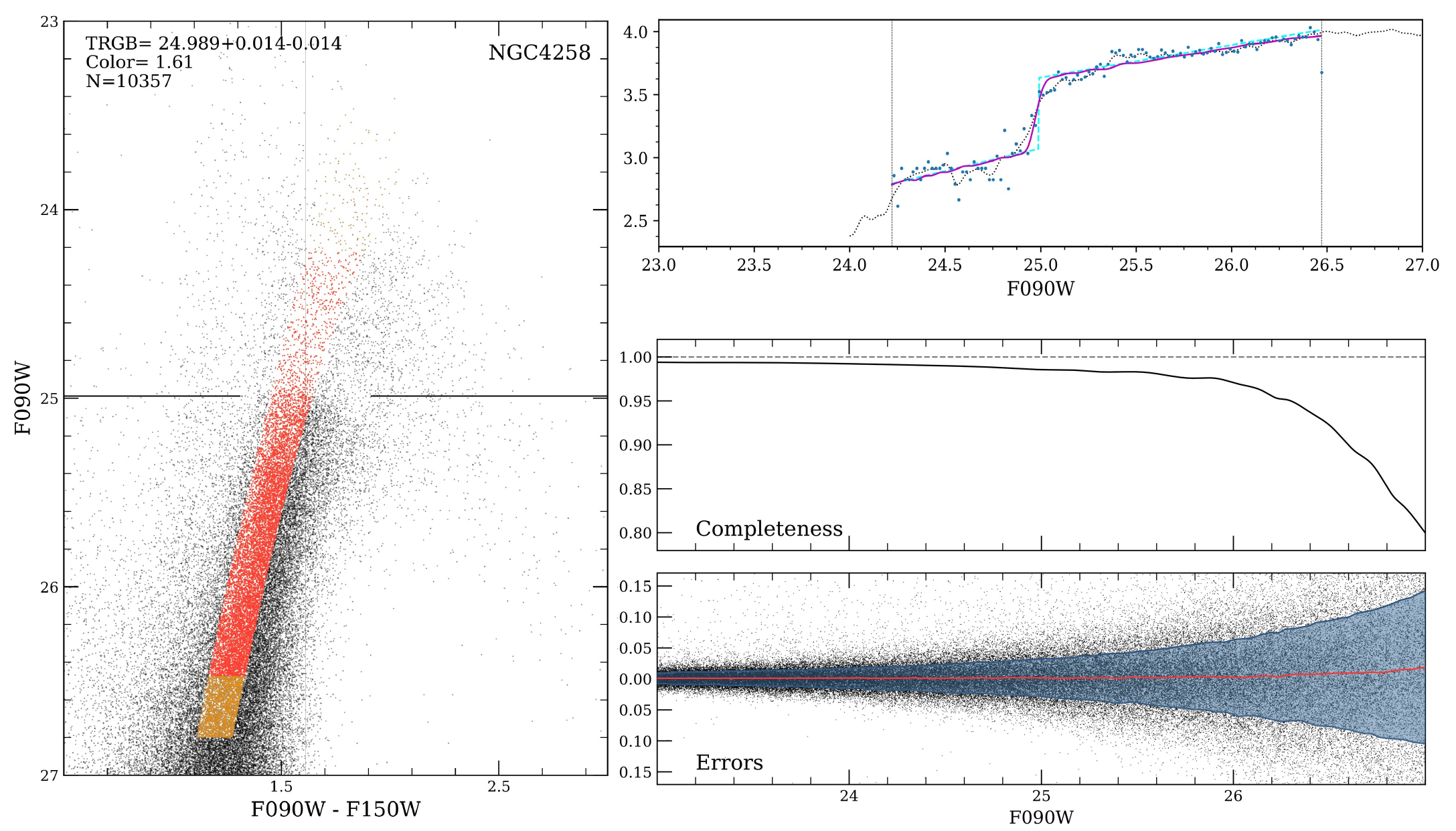}
    \caption{A thick slice of the RGB in the case of the galaxy NGC\,4258. In separate panels are the corresponding LF (top) and completeness and error dependencies with F090W magnitude (bottom).
    }
    \label{fig:NGC4258:slice:thick}
\end{figure}

The two panels of Figure~\ref{fig:NGC4258:slice:thin} illustrate examples of thin slicing of the RGB.  
For the TRGB zero point and color relation calibration it is desired to detect the TRGB with an uncertainty not exceeding about 0.02 mag (or 1$\%$ in distance). 
Unfortunately, the number statistics do not allow for very thin slices.
It was possible to select only 2--4 independent slices in the blue flat part of the TRGB in the case of NGC\,4258.
Moreover, the bluest slice suffers from small number statistics, and the boundary slice between the flat and inclined parts may already be subject to the color dependence effect.
Thus, we made the choice to use overlapping slices, which effectively smooths the running average TRGB values.
As a consequence, the values are no longer statistically independent, but the resultant curve describes the TRGB behavior in more robust detail.

The limiting case of increasing the slice thickness is to use a slice that covers almost the entire flat part of the TRGB. Figure~\ref{fig:NGC4258:slice:thick} illustrates the star selection and the maximum-likelihood fit to the resulting LF for such a thick slice in the galaxy NGC\,4258, encompassing the color range $1.45 \leq (\mathrm{F090W}-\mathrm{F150W})_0 \leq 1.65$ at the tip.

\begin{figure}
    \centering
    \includegraphics[width=1.0\linewidth]{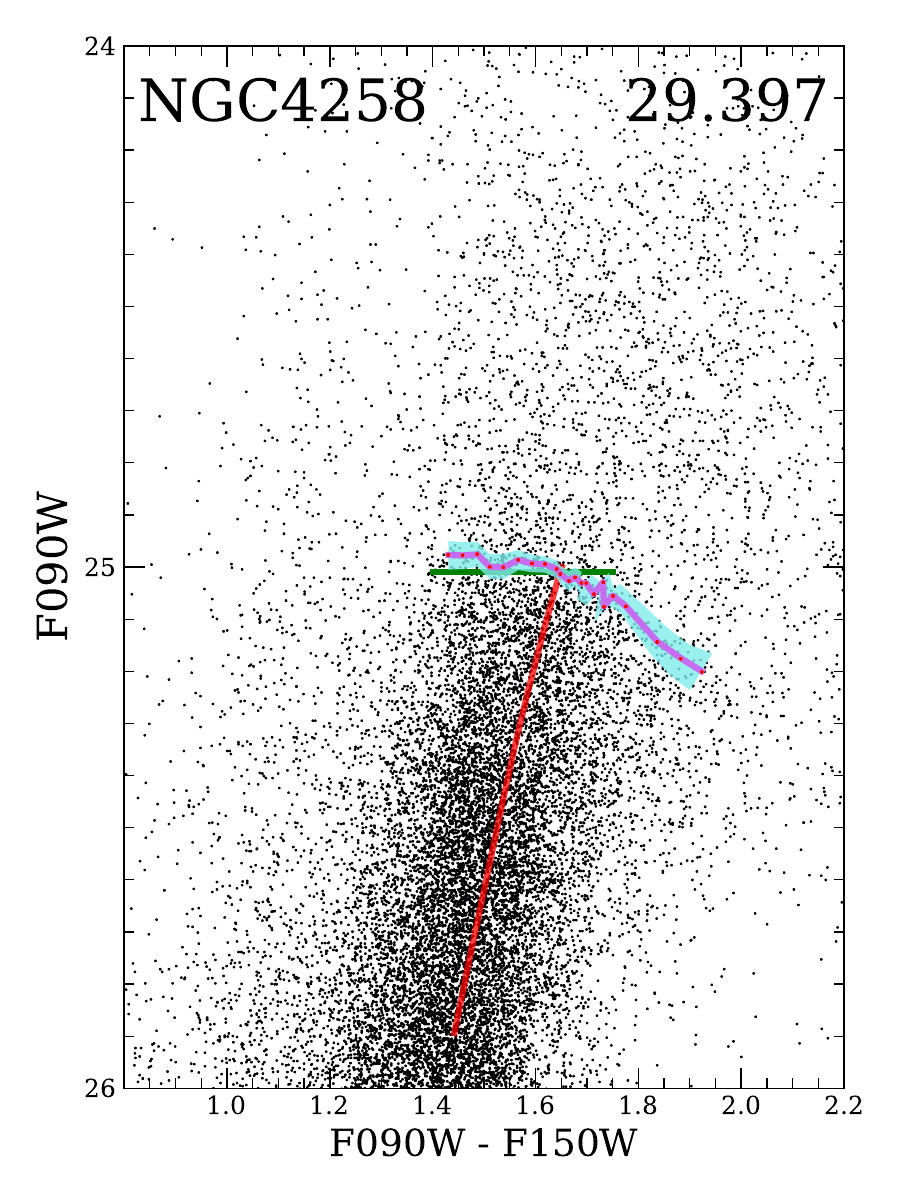}
    \caption{TRGB fit as a function of color for the galaxy NGC\,4258. The running mean of slices as a function of color are shown by the magenta line with cyan band uncertainties. The red band is the RGB isochrone for a 10~Gyr population with $[M/H] = -0.57$. The green constant value line at $\mathrm{F090W}_0=25.01$ corresponds to the fit by \citet{2024ApJ...966...89A} with revised photometry and after adjustment for reddening.
    }
    \label{fig:NGC4258:CMD:TRGB}
\end{figure}

Figure~\ref{fig:NGC4258:CMD:TRGB} shows $\mathrm{F090W}_0$ versus $(\mathrm{F090W}-\mathrm{F150W})_0$ CMD for NGC\,4258 with superposition of the color dependent results from RGB slicing.
The fit over the range 1.30--1.75 from \citet{2024ApJ...966...89A} is shown as a straight line. 
The constraints on the TRGB redward of $(\mathrm{F090W}-\mathrm{F150W})_0 = 1.7$ are tentative with this data for NGC\,4258.
The constraints in this red regime are much stronger from the CMDs of the elliptical galaxies which contain more high-metallicity stars, as discussed in the next secion.

\section{TRGB Color relation}
\label{sec:colorrel}

\begin{figure*}[h]
    \centering
    \includegraphics[width=1\linewidth]{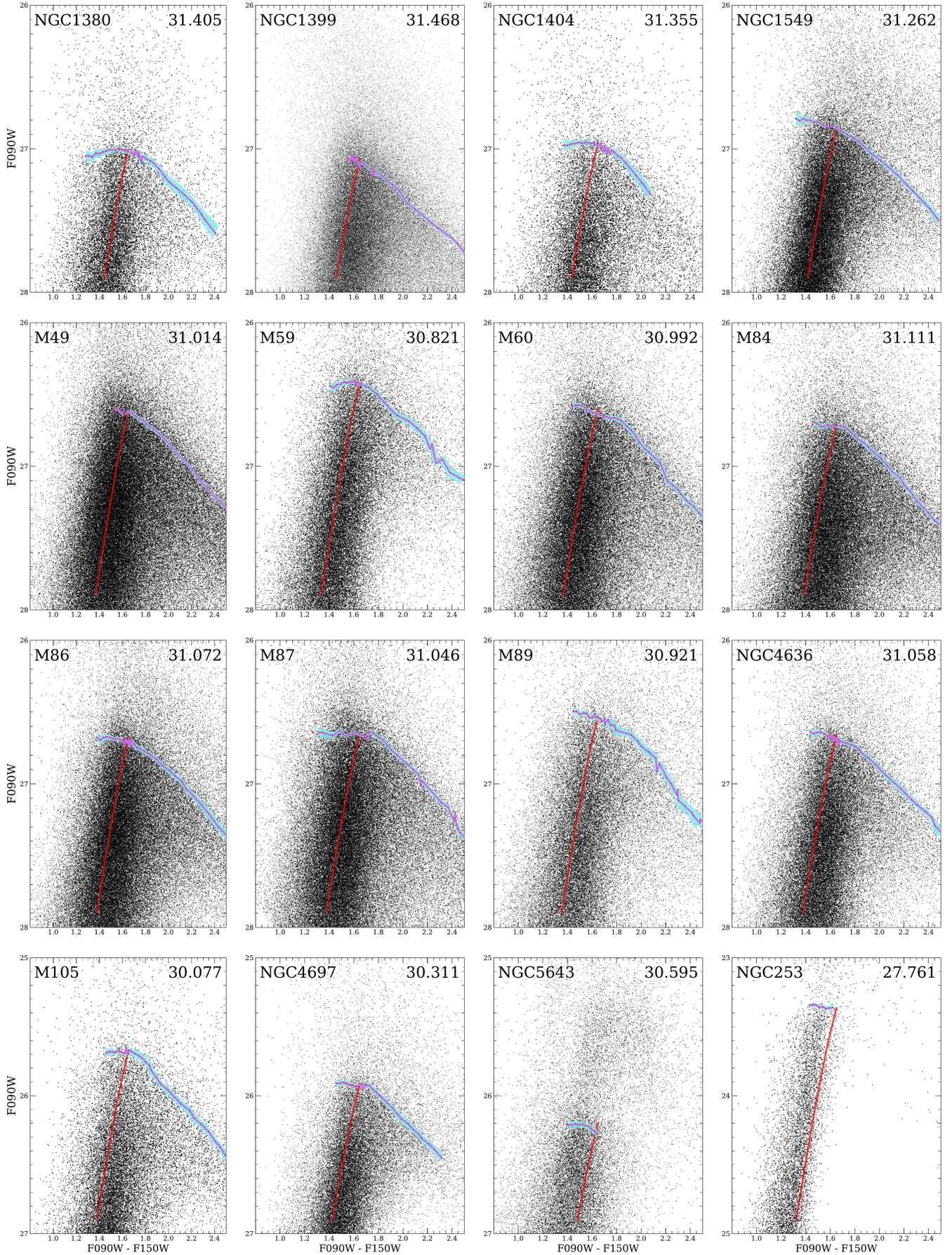}
    \caption{Color relations for all the galaxies except NGC\,4258. As in Figure~\ref{fig:NGC4258:CMD:TRGB}, cyan curves illustrate the TRGB limits as a function of $(\mathrm{F090W}-\mathrm{F150W})_0$ colors. The isochrone for 10 Gyr that ends where the TRGB color slope changes is shown in red.}
    \label{fig:relations-collage}
\end{figure*}

In total, 17 galaxies were used to establish the TRGB color-magnitude relation, including 10 galaxies from the Virgo Cluster and its surrounding environment, three within the Fornax Cluster, as well as NGC\,253, NGC\,1549, NGC\,4258 and NGC\,5643. The resulting TRGB color-magnitude fits for all galaxies except NGC\,4258 are superimposed on CMD in Figure~\ref{fig:relations-collage}. The case of NGC\,4258 is shown in Figure~\ref{fig:NGC4258:CMD:TRGB}. 

The photometric quality of the data varies from galaxy to galaxy which influences the precision of the TRGB color-magnitude relations. In some cases it is difficult to construct a high-precision TRGB color relation due to insufficient statistics in individual slices. Consequently, two samples are adopted for the following analysis: the first comprises galaxies with minimal photometric scatter between slices at the TRGB defining precise color-magnitude relations, while the second includes the full sample. The uncertainties are smallest for bright RGB stars in the F150W filter. A total of 12 galaxies have been selected for the final color-magnitude relation construction (identified in Table~\ref{tab:distances}). The resulting dependencies show excellent agreement for red, high-metallicity stars, even though the quality of the maximum-likelihood fitting in individual cases may be suboptimal.

\begin{figure}
    \centering
    \includegraphics[width=1\linewidth]{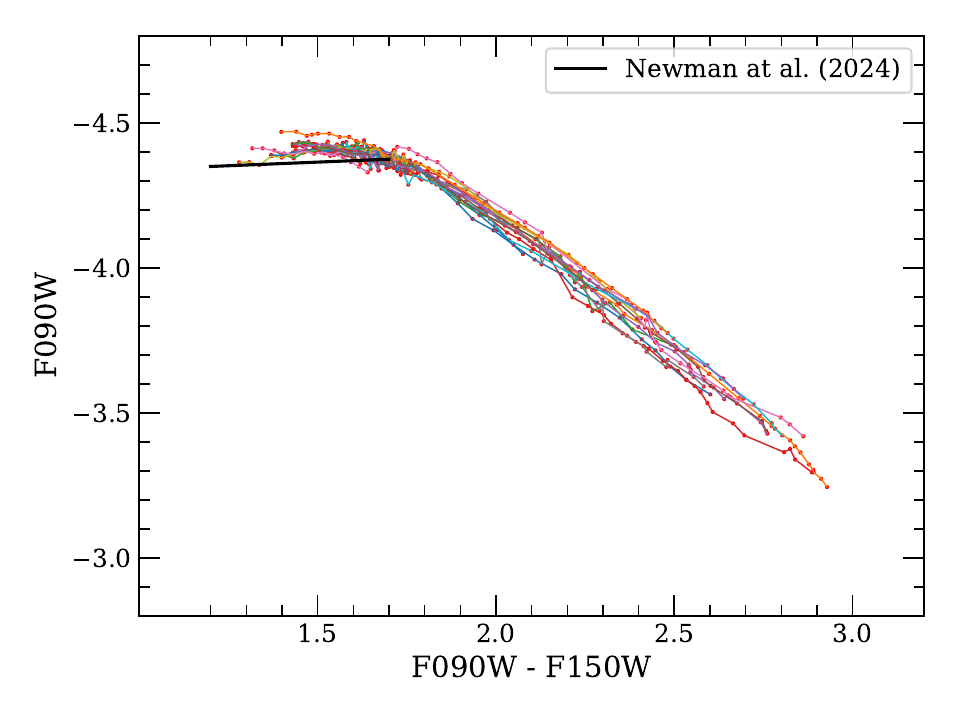}
    \includegraphics[width=1\linewidth]{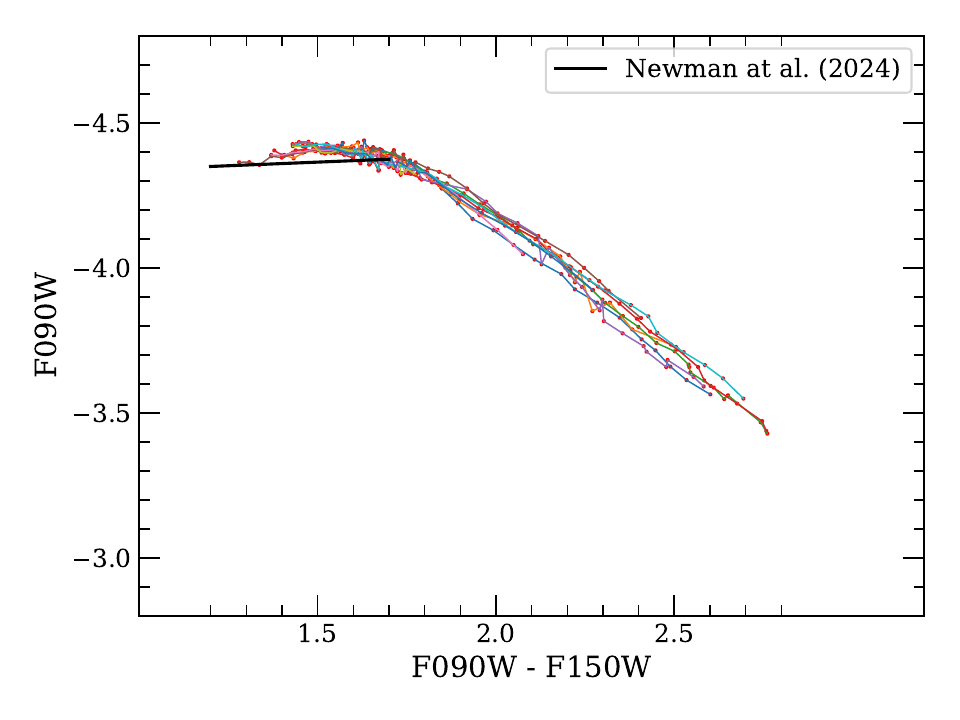}
    \caption{Top panel: Resulting TRGB color relations for the entire sample. Bottom Panel: Resulting TRGB color relations for galaxies with high photometry quality and without significant fit problems.}
    \label{fig:fitted:all}
\end{figure}

To construct a reliable color calibration, we need to combine data from all sample galaxies, taking care to account for color shifts due to reddening. 
The relations were aligned in brightness by fitting relative shifts with respect to NGC\,4258, which was chosen as the reference point for the color-dependence relation. Using the points in the same color interval as we have for NGC\,4258 we can calculate the weighted shift required for the alignment:
\begin{equation}
    \delta = \left( \int \frac{f(x) - g(x)}{\sigma_f^2(x) + \sigma_g^2(x)} dx \right) \cdot \left( \int \frac{dx}{\sigma_f^2(x) + \sigma_g^2(x)} \right)^{-1}
\end{equation}
where $f(x)$ denotes the TRGB value at color $x$ for NGC\,4258 and $g(x)$ denotes the equivalent value for another galaxy in the sample.
This approach allows us to minimize the systematic shift between relations over a common color range. For the high-quality photometry sample, this approach achieves excellent agreement for the entire relation, as shown in the bottom of Figure~\ref{fig:fitted:all}.
However, the top panel of Figure~\ref{fig:fitted:all} demonstrates that when all galaxies are included, the blue part of the relation exhibits significantly larger scatter. A comparison between the full set of galaxies and the subset with high-quality photometry reveals that galaxies with precise photometry display a much smaller spread, particularly in the critical region of the flat blue part of the TRGB.

\begin{figure}
    \centering
    \includegraphics[width=1\linewidth]{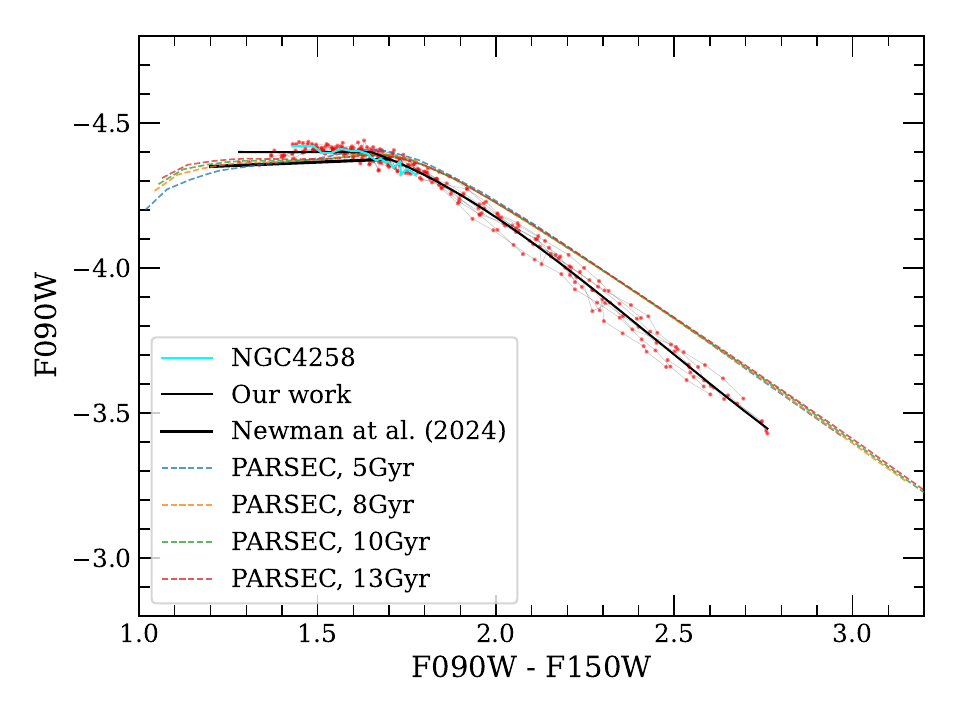}
    \caption{Color relations for high photometry quality galaxies. The \citet{2024ApJ...975..195N} calibration blueward of $(\mathrm{F090W}-\mathrm{F150W})_0=1.7$ is shown in the thick black line. Theoretical PARSEC expectations redward of $(\mathrm{F090W}-\mathrm{F150W})_0=1.3$ are shown by dotted lines.}
    \label{fig:relations}
\end{figure}

As a result, it was decided to use only high-quality photometric galaxies to construct the definitive TRGB color relation. The data were fitted using a fourth-degree polynomial, and the resulting TRGB color relation is illustrated in Figure~\ref{fig:relations}.
The final TRGB color magnitude relation is described by the piecewise function:
\begin{equation}
M^\mathrm{TRGB}_\mathrm{F090W} =
\begin{cases}
    -4.398 & \text{if } x < 0 \\
    A + B x + C x^2 + D x^3  & \text{if } x \geq 0 \\
\end{cases}
\label{eq:clt0}
\end{equation}
where $x = (\mathrm{F090W} - \mathrm{F150W})_0 - 1.65$ and the coefficients A, B, C, D are given in Table~\ref{tab:coeffs}.\textbf{}

\begin{table}[h]
    \centering
    \caption{Coefficients for derived color-magnitude relation.} 
    \label{tab:coeffs}
    \begin{tabular}{rrrr}
        \hline\hline
        \multicolumn{1}{c}{$A$} & \multicolumn{1}{c}{$B$} & \multicolumn{1}{c}{$C$} & \multicolumn{1}{c}{$D$} \\
        \hline
        $-4.398$ & $0.431$ & $0.712$ & $-0.293$ \\
        $\pm 0.017$ & $\pm 0.027$ & $\pm 0.074$ & $\pm 0.051$ \\
        \hline\hline
    \end{tabular}
\end{table}

The isochrone that passes through the point of the relation bend has $[M/H] = -0.49$ for 7 Gyr or $[M/H] = -0.57$ for 10 Gyr.

\section{Distance measurements}

\begin{table*}
\centering
\caption{New and literature distance moduli. 
Distance moduli are given for both approaches used in this work: applying MLE method directly to ``thick" slices and for ``thin slices''. Average differences between values in current and previous works  0.037~mag in the thick slice case and 0.028~mag in the thin slice case. The uncertainty on the distance moduli do not include the systematic uncertainty in the NGC\,4258 zero-point scaling of 0.046~mag.
}
\label{tab:distances}
\begin{threeparttable}
    \begin{tabular}{l@{}cccccc}
        \hline\hline
        Name        & & $A_\mathrm{F090W}$   & $m_\mathrm{TRGB}^\mathit{thick}$               & $(m-M)_0^\mathit{thick}$                & $(m-M)_0^\mathit{thin}$         & Literature \\
        \hline
        NGC\,1380    & \tnote{*} & 0.021 & 27.007 $\pm$ 0.013 & 31.405 $\pm$ 0.027 & 31.408 $\pm$ 0.025 & 31.397 $\pm$ 0.066 \tnote{a} \\
        NGC\,1399    &           & 0.015 & 27.070 $\pm$ 0.012 & 31.468 $\pm$ 0.026 & 31.498 $\pm$ 0.026 & 31.511 $\pm$ 0.068 \tnote{a} \\
        NGC\,1404    & \tnote{*} & 0.014 & 26.957 $\pm$ 0.014 & 31.355 $\pm$ 0.027 & 31.366 $\pm$ 0.025 & 31.364 $\pm$ 0.067 \tnote{a} \\
        \hline
        M\,49        &           & 0.027 & 26.616 $\pm$ 0.012 & 31.014 $\pm$ 0.026 & 31.016 $\pm$ 0.027 & 31.091 $\pm$ 0.071 \tnote{b} \\
        M\,59        & \tnote{*} & 0.039 & 26.423 $\pm$ 0.014 & 30.821 $\pm$ 0.028 & 30.821 $\pm$ 0.026 & 30.841 $\pm$ 0.074 \tnote{b} \\
        M\,60        &           & 0.032 & 26.594 $\pm$ 0.012 & 30.992 $\pm$ 0.026 & 31.009 $\pm$ 0.025 & 31.061 $\pm$ 0.072 \tnote{b} \\
        M\,84        & \tnote{*} & 0.049 & 26.713 $\pm$ 0.015 & 31.111 $\pm$ 0.028 & 31.112 $\pm$ 0.027 & 31.195 $\pm$ 0.075 \tnote{b} \\
        M\,86        & \tnote{*} & 0.036 & 26.674 $\pm$ 0.014 & 31.072 $\pm$ 0.027 & 31.096 $\pm$ 0.025 & 31.139 $\pm$ 0.072 \tnote{b} \\
        M\,87        &           & 0.028 & 26.648 $\pm$ 0.013 & 31.046 $\pm$ 0.027 & 31.046 $\pm$ 0.025 & 31.062 $\pm$ 0.071 \tnote{b} \\
        M\,89        & \tnote{*} & 0.050 & 26.523 $\pm$ 0.016 & 30.921 $\pm$ 0.029 & 30.923 $\pm$ 0.028 & 30.933 $\pm$ 0.075 \tnote{b} \\
        M\,105       & \tnote{*} & 0.030 & 25.679 $\pm$ 0.019 & 30.077 $\pm$ 0.030 & 30.078 $\pm$ 0.028 & 30.085 $\pm$ 0.076 \tnote{b} \\
        NGC\,4636    & \tnote{*} & 0.035 & 26.660 $\pm$ 0.013 & 31.058 $\pm$ 0.027 & 31.070 $\pm$ 0.026 & 31.120 $\pm$ 0.072 \tnote{b} \\
        NGC\,4697    & \tnote{*} & 0.036 & 25.913 $\pm$ 0.015 & 30.311 $\pm$ 0.028 & 30.324 $\pm$ 0.028 & 30.330 $\pm$ 0.073 \tnote{b} \\
        \hline
        NGC\,253     & \tnote{*} & 0.023 & 23.363 $\pm$ 0.015 & 27.761 $\pm$ 0.028 & 27.770 $\pm$ 0.036 & ---    \\
        NGC\,1549    &           & 0.015 & 26.864 $\pm$ 0.012 & 31.262 $\pm$ 0.026 & 31.231 $\pm$ 0.026 & ---    \\
        NGC\,5643    & \tnote{*} & 0.228 & 26.197 $\pm$ 0.016 & 30.595 $\pm$ 0.046 & 30.621 $\pm$ 0.052 & 30.57 $\pm$ 0.06  \tnote{c} \\
        \hline
        \textbf{NGC\,4258} & \tnote{*} & \textbf{0.020} & \textbf{24.989 $\pm$ 0.014} & \textbf{29.387 $\pm$ 0.027 } & \textbf{29.397 $\pm$ 0.026 } & \textbf{29.397  $\pm$ 0.032} \tnote{d} \\

        \hline\hline
    \end{tabular}
\begin{tablenotes}
\item[*] Galaxies contributing to the high quality sample represented in bottom panel Figure~\ref{fig:fitted:all}. 
\item[a] \citet{2024ApJ...973...83A}, includes NGC\,4258 zero-point uncertainty
\item[b] \citet{2025ApJ...982...26A}, includes NGC\,4258 zero-point uncertainty
\item[c] The modulus for NGC\,5643 \citep{2024ApJ...976..177L} results from a Sobel filter analysis
\item[d] Geometric maser distance modulus \citep{2019ApJ...886L..27R}
\end{tablenotes}
\end{threeparttable}
\end{table*}

In Table~\ref{tab:distances}, distance moduli based on JWST observations are reported for 13 galaxies in the Fornax and Virgo cluster regions \citep{2024ApJ...973...83A, 2025ApJ...982...26A} as well as for WLM \citep{2024ApJ...961...16M}, NGC\,5643 \citep{2024ApJ...976..177L}, and NGC\,1549 \citep{2023jwst.prop.3055T}.
The first 13 results were derived from a similar MLE methodology used in the current paper with the difference of a red color cut at $(\mathrm{F090W}-\mathrm{F150W})_0 = 1.75$ (blue color cuts do not significantly affect results) and a simple vertical color cut instead of slicing.
These published distance moduli were scaled to fits to NGC\,4258 at the maser modulus 29.397~mag \citep{2019ApJ...886L..27R, 2024ApJ...966...89A}.
In the case of NGC\,5643 the literature modulus \citep{2024ApJ...976..177L} was derived from a Sobel filter which is not directly comparable.

Table~\ref{tab:distances} reports two moduli for each galaxy based on the methods described in this paper.
In the cases of the moduli with the superscript $\mathit{thin}$, results are derived by the thin slice method with fits to the mean relation displayed in Figure~\ref{fig:relations} that are, in turn, normalized to NGC\,4258.
In the cases of moduli with superscript $\mathit{thick}$, MLE fits are made using a single wide slice limited on the red side at the RGB isochrone reaching $(\mathrm{F090W}-\mathrm{F150W})_0 = 1.65$ at the TRGB, the color break-point (see the red lines in the panels of Figure~\ref{fig:relations-collage}).
Since the TRGB measurement and the red edge of the slice are coupled, a slight iteration is required for the optimal fit. This MLE fit is given in the table column $m_{TRGB}^\mathit{thick}$.  
Note again that adjustments for reddening are made {\it before} the MLE analysis is carried out to assure the proper location of the 1.65 mag color break-point. 
The distance modulus is then $\mu = m_\mathrm{TRGB} + 4.398$ as given by Eq.~\ref{eq:clt0}.

\begin{figure}
    \centering
    \includegraphics[width=0.9\linewidth]{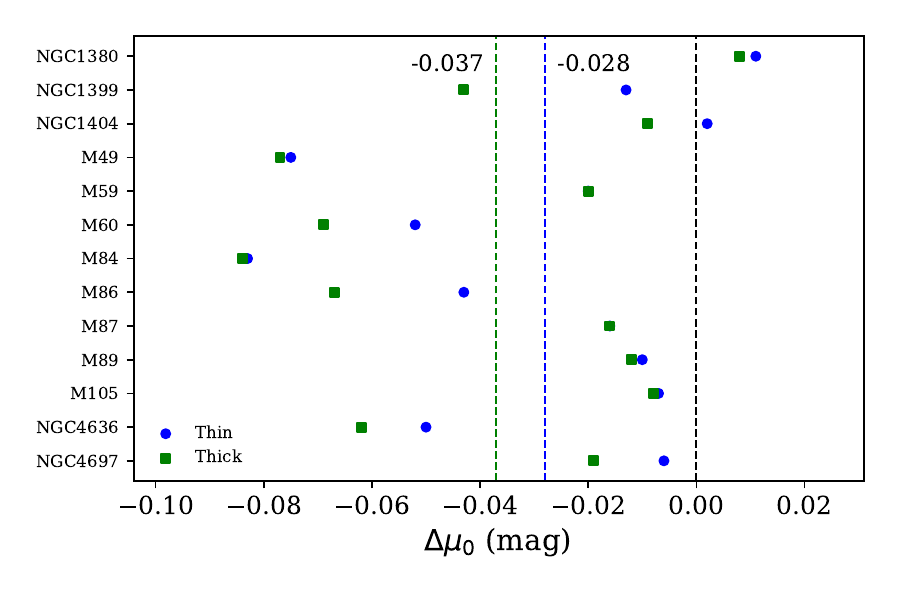}
    \caption{Offsets between separately reported TRGB magnitudes.
    The shifts are defined as $\Delta\mu_0 = m_{\mathrm{thin/thick}} - m_{\mathrm{literature}}$.
Blue and green symbols denote the thin- and thick-slice measurements, respectively.
The vertical dashed black line indicates zero offset, while the dotted blue and green lines show the mean shifts for the thin- and thick-slice approaches.
These systematic shifts are partially driven by updates to the JWST pipeline and DOLPHOT.}
    \label{fig:shifts}
\end{figure}

Differences between the literature moduli and each of the two moduli found here are shown graphically in Figure~\ref{fig:shifts}.
With both methodologies, there are small but systematic shifts to lower values.
Averaging over the top 13 galaxies in the table, there is a shift of $-0.037\pm0.009$~mag between the new thick slice moduli and the literature and a shift of $-0.028\pm0.008$~mag between the thin slice fits defined over the color range 1.4 to 1.8 mag and the literature.

The determination of our uncertainties is as follows. For the statistical uncertainty on the distance moduli, we add in quadrature 
1) the uncertainty in the measurement of the tip itself (shown in Table \ref{tab:distances} in the $m_{TRGB}^\mathit{thick}$ column), 
2) the uncertainty in the foreground reddening, defined as the quadrature sum of typical 16\% error~\citep{1998ApJ...500..525S}, $0.16 \times A_\mathrm{F090W}$, and the typical spread in the regions of the small extinction of $0.003\times E(B-V)$~\citep{2010ApJ...719..415P} corresponding to 0.005~mag in the F090W band, 
3) 0.02~mag to account for uncertainties within DOLPHOT's aperture correction procedure, 
and 4) 0.01~mag to cover the PSF stability of JWST/NIRCam. 
This overall term is provided within Table \ref{tab:distances} as the uncertainty on the distance moduli.

To determine the systematic uncertainty arising from the zero-point calibration, we add in quadrature 1) the overall statistical uncertainty on the measurement of the TRGB in NGC\,4258 (0.027~mag), 2) the uncertainty in its geometric distance (\citealt{2019ApJ...886L..27R}; 0.032~mag), and 3) 0.02~mag to account for any intrinsic stellar population variation in the magnitude of the TRGB \citep{2019ApJ...880...63M, 2024ApJ...966...89A}, summing to $\pm0.047$~mag.

\section{Discussion}
\label{sec:dicuss}

The distance moduli found for our primary sample of 13 galaxies listed in Table~\ref{tab:distances}, for two MLE methodologies analyzing the same data, differ from each other by a statistically insignificant $0.007$~mag.
As a practical matter, we give preference to the thick slice procedure that limits colors on red side at the juncture of the RGB isochrone with the TRGB at $(\mathrm{F090W}-\mathrm{F150W})_0=1.65$~mag.
We prefer the thick slice fits because the RGB of lower metallicity populations in the halos of most galaxies fall on the blue of the 1.65~mag color break.

With either thick or thin slice procedure, measured distance moduli  are slightly lower than found analyzing the same data with related techniques by \citet{2024ApJ...966...89A, 2024ApJ...973...83A, 2025ApJ...982...26A} that did not take into account the slope of the TRGB at redder colors. In those earlier published cases, the MLE methodology assumed a flat TRGB extending to $(\mathrm{F090W}-\mathrm{F150W})_0=1.75$~mag.
Hence, RGB stars were being included redward of the 1.65~mag downturn in the TRGB with the consequence that the averaged TRGB magnitude assuming a flat value were reduced. Differences in reported distances are minimized given the related adjustment required for the calibrator galaxy NGC\,4258.
Differences with the 13 galaxies contributing to Figure~\ref{fig:shifts} persist because the galaxies in all these cases are early types with more substantial populations near the color break than is the case with NGC\,4258.
With the mean of the thick and thin fits, the average shift in
modulus from the earlier papers I and II in this series \citep{2024ApJ...973...83A, 2025ApJ...982...26A} is $-0.032$~mag or a reduction of 1.5\% in distances.
An immediate consequence is a shift of the absolute calibration of SBF distances resulting in a slight increase in the $H_0$ value reported in series Paper~III  \citep{2025ApJ...987...87J}.

The other comparison that can be made with JWST observations at $(\mathrm{F090W}-\mathrm{F150W})$ is with \citet{2024ApJ...975..195N}.
That work found the TRGB to be essentially flat extending to $(\mathrm{F090W}-\mathrm{F150W})_0=1.68$, consistent with the current study.
However there is a difference in absolute calibration.
\citet{2024ApJ...975..195N} find $M^\mathrm{TRGB}_\mathrm{F090W} = -4.36$~mag\footnote{The \citet{2024ApJ...975..195N} magnitudes in the Vega-Sirius system are translated here to the Vega system used in this paper.} based on the HST calibration of the TRGB at F814W by \citet{2021ApJ...919...16F} and distances to the two Local Group dwarfs WLM and Sextans~A.
This calibration is to be compared with our value of $M^\mathrm{TRGB}_\mathrm{F090W} = -4.40$~mag (Eq.~\ref{eq:clt0}) from the maser distance to NGC\,4258.
This difference of 0.04~mag, almost 2\% in distance, is larger than the relative uncertainties in JWST TRGB distance estimates at F090W and point to the ongoing need to address the absolute scaling.



\section{Conclusion}
\label{sec:conclude}

We have developed a new TRGB measurement procedure and code written in Python and .NET, utilizing an earlier method for TRGB MLE calculations with minor improvements. A novel methodology was developed which incorporates statistically robust and physically motivated calculations for selecting RGB star subsamples for MLE analysis, enabling us to measure the color-dependence of the TRGB for higher metallicities and redder colors with JWST at F090W.

All analyzed galaxies exhibit good agreement between their individual TRGB relations and the generalized relation, with uncertainties of 0.023 mag for a high-precision photometry sample and 0.029 mag for the full sample of galaxies. These results enable a more precise application of the original TRGB methodology to galaxies containing high-metallicity stars, significantly enhancing its utility for cosmological distance measurements.

The TRGB color-magnitude relation has been derived at $M^\mathrm{TRGB}_\mathrm{F090W}$ as a function of $(\mathrm{F090W}-\mathrm{F150W})_0$ color
based on JWST observations of seventeen galaxies, primarily located in the Virgo and Fornax clusters. 
Within current measurement capabilities, the relation is flat at $M^\mathrm{TRGB}_\mathrm{F090W} = -4.398$~mag blueward of $(\mathrm{F090W}-\mathrm{F150W})_0=1.65$, then breaks to fainter magnitudes in the F090W band at redder colors.
From theoretical isochrones, the break occurs at a metallicity of $[M/H] = -0.57$ for stars of age 10~Gyr.
There is good agreement between the empirical and theoretical decrease of the $M^\mathrm{TRGB}_\mathrm{F090W}$ magnitude with increasing metallicity beyond the break, although the empirical fall-off is slightly faster than the model.

The empirical relation was determined by combining broken-line fits across the color break at $(\mathrm{F090W}-\mathrm{F150W})_0=1.65$ for individual galaxies.
The integration of 16 individual galaxies with the absolute calibration provided by NGC 4258 allowed us to estimate relative distances and construct a generalized relation for the absolute TRGB magnitude as a function of color in the range $(\mathrm{F090W}-\mathrm{F150W})_0 = [1.4, 3.0]$.

\begin{acknowledgments}

The careful attention and useful  comments by the anonymous referee are most appreciated.

This work is based on observations made with the NASA/ESA/CSA JWST, program 3055. The data were obtained from the Mikulski Archive for Space Telescopes at the Space Telescope Science Institute, which is operated by the Association of Universities for Research in Astronomy, Inc., under NASA contract NAS 5-03127 for JWST. The specific observations analyzed can be accessed via \href{https://archive.stsci.edu/doi/resolve/resolve.html?doi=10.17909/20tm-sg94}{10.17909/20tm-sg94}.

This research made use of the NASA Astrophysics Data System and the NASA/IPAC Extragalactic Database (NED) funded by the National Aeronautics and Space Administration and operated by the California Institute of Technology.

MIC, DIM and LNM are supported by the Russian Science Foundation grant \textnumero~24--12--00277. 
RBT, GSA and YC acknowledge financial support from JWST GO--3055.
MC acknowledges support from ASI–INAF grant no. 2024-10-HH.0 (WP8420), the ESO Scientific Visitor Programme, and INAF GO-grant no. 12/2024 (P.I. M. Cantiello). RH and MC acknowledge support from the project “INAF-EDGE” (Large Grant 12-2022, P.I. L. Hunt).

\end{acknowledgments}

\newpage

\appendix

\section{Segmentation of the RGB}
\label{app:RGBsegmentation}

The strong color dependence observed in the TRGB for $(\mathrm{F090W} - \mathrm{F150W})_0 > 1.65$ necessitates an accurate characterization of the color distribution. To achieve this, a multi-stage process to isolate distinct regions along the RGB is employed.

Initially, the RGB behavior is described using a broken line, where the segment corresponding to blue colors is approximated as a constant, and the segment corresponding to redder colors roughly follows the TRGB color relation. This relation is represented by the blue lines in Figure~\ref{fig:strips}. Each line is parameterized by the constant value for the blue segment and the inclination angle for the red segment. The first two upper broken lines are defined by the globally fitted TRGB value for the galaxy and an inclination angle of 45 degrees. Subsequently, the proportion of blue and red stars within a horizontal strip are estimated, with the separation defined at a color value of 1.75. This ratio is determined individually for each galaxy and on average it was within 0.4-0.7 for most galaxies.

Maintaining the observed blue-to-red star ratio along the RGB, the color boundary and the inclination angle of the red segment is iteratively adjusted to define additional lines. These lines enable the division of the RGB into evenly spaced strips originating from the TRGB. Within each strip, colors are computed corresponding to specific increments of the color distribution.

The color-magnitude relation for the derived increments are then fitted using a second-order polynomial, as illustrated by the magenta lines in Figure~\ref{fig:strips}. These polynomial relations define slices that isolate distinct stellar populations along the RGB.
Finally, for each slice, the magnitude distribution is fit using a broken exponential RGB LF. The maximum-likelihood approach is then applied to detect the TRGB.

\begin{figure*}
    \centering
    \includegraphics[width=0.45\linewidth]{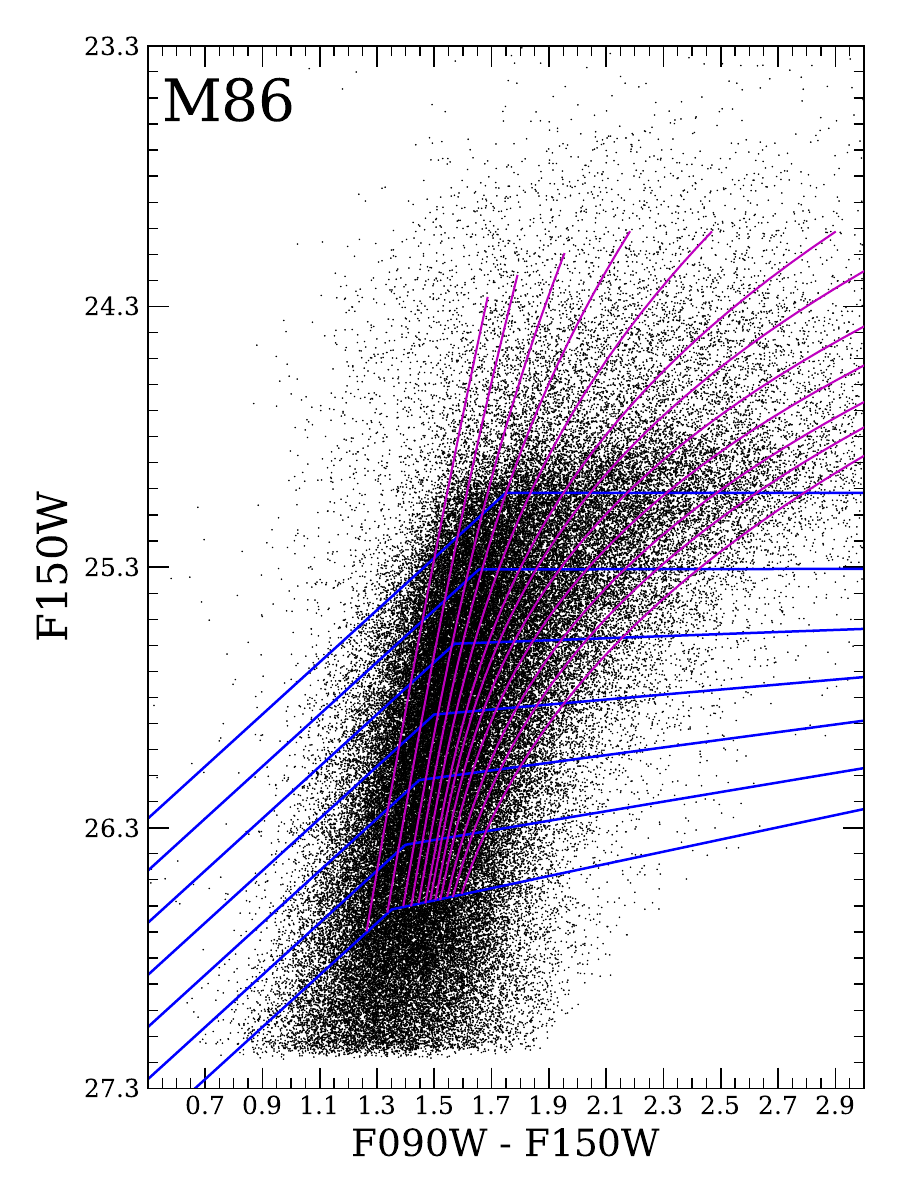}
    \includegraphics[width=0.45\linewidth]{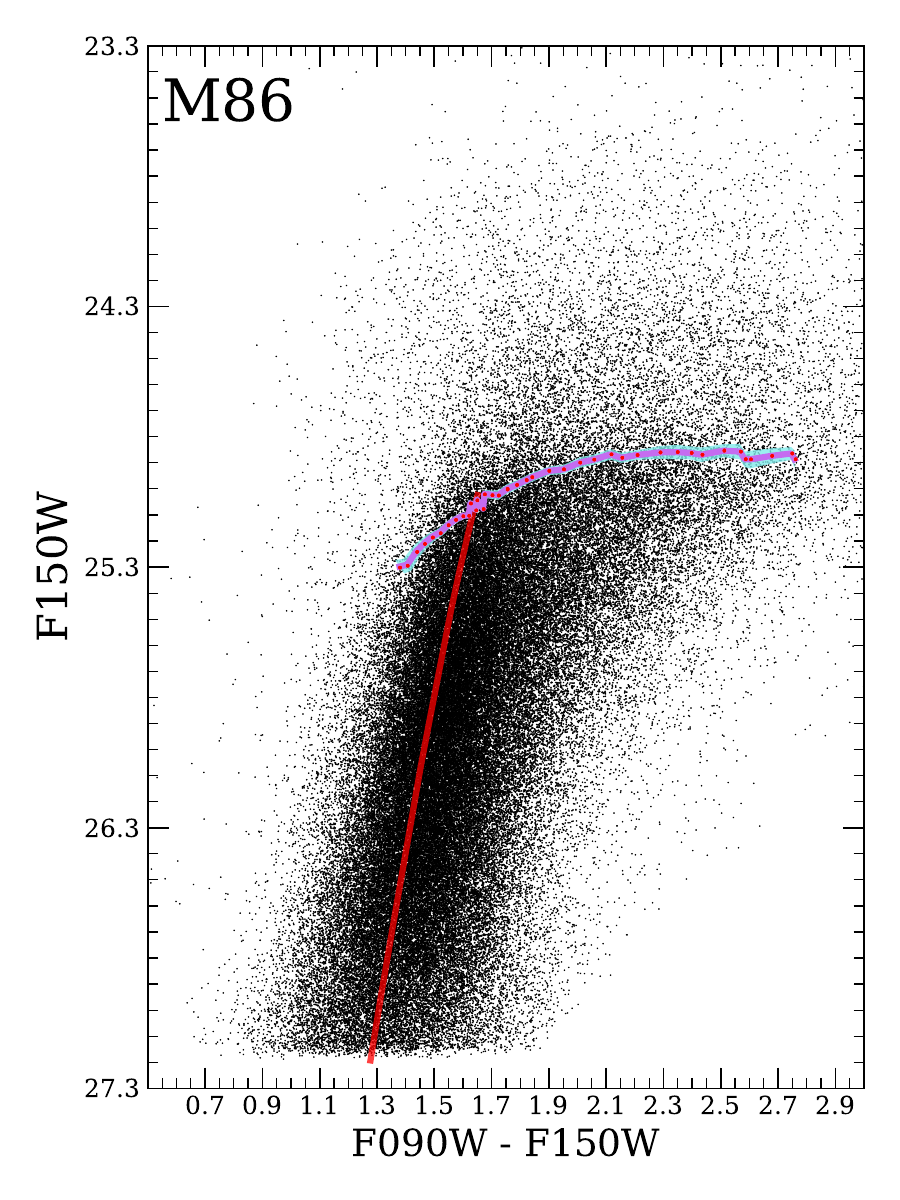}
    \caption{CMD with $\mathrm{F150W}_0$ magnitudes versus $(\mathrm{F090W}-\mathrm{F150W})_0$ color for the M\,86 galaxy. Left panel: segmentation applied for TRGB thin-slice fitting in the reddest and bluest regions of the RGB, based on theoretical isochrones;
    Right panel: the TRGB fit obtained from thin-slice segmentation.
    The isochrone reaching the TRGB at F090W$-$F150W=1.65 is shown in red.}
    \label{fig:F150W}
\end{figure*}

While this segmentation approach performs well in the densely populated regions of the RGB, it becomes challenging to apply the same method near the branch edges. To address this, we employ theoretical isochrones to achieve comparable slicing quality. This enables us to extend the color-magnitude relation into the bluer part of the RGB, down to a color of $(\mathrm{F090W} - \mathrm{F150W})_0\approx1.45$ in several galaxies, as well as into the redder RGB populations.
The case of slicing based on theoretical isochrones is demonstrated in the left panel of Figure~\ref{fig:F150W}.
However, the corresponding data points carry significant uncertainties due to limited statistics.
Because photometric errors and biases cannot be tracked continuously along the RGB slice, appropriately selected artificial-star statistics were applied. 
For the completeness correction, the mean color for each slice  was computed, allowing the slice completeness to be approximated as the product of the completeness functions in the two filters. This approach enables the use of quasi-two-dimensional statistics while requiring only a minimal number of artificial stars.

\section{Maximum Likelihood fits at red colors}
\label{app:MLfit}

Inspection of the CMD in Figure~\ref{fig:relations-collage} makes clear that the TRGB magnitude at F090W falls off redward of $(\mathrm{F090W}-\mathrm{F150W})_0 \sim 1.65$.
Equally, at the halo locations optimal for TRGB measurements the abundances of RGB stars are falling off toward redder colors.
The sloping color dependence and diminished statistics within slices of the RGB lead to substantial uncertainties in F090W TRGB measurements at red colors.

The right panel of Figure~\ref{fig:F150W} presents the CMD for M\,86, analogous to that shown in Figure~\ref{fig:relations-collage}, but now with the F150W filter on the vertical axis.
The locus of the TRGB as traced by the cyan curve increases slightly instead of getting fainter.
Whereas with F090W magnitudes the TRGB is roughly flat at blue colors and falling off at red colors, at F150W the TRGB is sloping upward at blue colors and becomes roughly flat at red colors.
The upward slope at blue colors with F150W has been demonstrated by \citet{2024ApJ...975..195N} although their sample does not include CMD that explore the flattening at high metallicities.

The consequence is that the TRGB is more robustly determined in the redward RGB slices with F150W magnitudes.
Not only is the CMD roughly flat but F150W magnitudes are brighter than F090W magnitudes and measured with higher accuracy.
Given these considerations, the TRGB in slices redward of approximately $(\mathrm{F090W}-\mathrm{F150W})_0 \simeq 1.7$ is measured in the F150W band. Fits for the color range $(\mathrm{F090W}-\mathrm{F150W})_0 \approx [1.65, 1.75]$ were performed in both the F090W and F150W filters.
The result is a saw-tooth pattern in the corresponding color–magnitude relations.
TRGB values fitted in F150W were transformed to the F090W band using a color defined as the mean of the slice color boundaries evaluated at the fitted TRGB magnitude level.

\newpage
\bibliographystyle{aasjournal}
\bibliography{paper}{}

\end{document}